\documentclass[11pt,prd,preprint,superscriptaddress]{revtex4}
 \pdfoutput=1
\usepackage{revsymb}
\usepackage{graphicx}
\usepackage{bm}
\usepackage{amsfonts}
\usepackage{amsmath}
\usepackage{amssymb}
\usepackage{graphicx}

\begin{document}

\title{Scalar-tensor extension of the $\Lambda$CDM model}

\date{\today}
\author{W.C. Algoner\footnote{E-mail:  w.algoner@cosmo-ufes.org}}
\affiliation{Universidade Federal do Esp\'{\i}rito Santo,
Departamento
de F\'{\i}sica\\
Av. Fernando Ferrari, 514, Campus de Goiabeiras, CEP 29075-910,
Vit\'oria, Esp\'{\i}rito Santo, Brazil}
\author{H.E.S. Velten\footnote{E-mail: velten@pq.cnpq.br}}
\affiliation{Universidade Federal do Esp\'{\i}rito Santo,
Departamento
de F\'{\i}sica\\
Av. Fernando Ferrari, 514, Campus de Goiabeiras, CEP 29075-910,
Vit\'oria, Esp\'{\i}rito Santo, Brazil}
\author{W. Zimdahl\footnote{E-mail: winfried.zimdahl@pq.cnpq.br}}
\affiliation{Universidade Federal do Esp\'{\i}rito Santo,
Departamento
de F\'{\i}sica\\
Av. Fernando Ferrari, 514, Campus de Goiabeiras, CEP 29075-910,
Vit\'oria, Esp\'{\i}rito Santo, Brazil}

\begin{abstract}
We construct a cosmological scalar-tensor-theory model in which the Brans-Dicke type scalar $\Phi$ enters the effective (Jordan-frame) Hubble rate as a simple modification of the Hubble rate of the $\Lambda$CDM model.
This allows us to quantify differences between the background dynamics of scalar-tensor theories and general relativity (GR) in a transparent and observationally testable manner in terms of one single parameter.
Problems of the mapping of the scalar-field degrees of freedom on an effective fluid description in a GR context are discused. Data from supernovae, the differential age of old galaxies and baryon acoustic oscillations are shown to
strongly limit potential deviations from the standard model.
\end{abstract}

\pacs{98.80.-k, 04.50.+h}

\maketitle
\date{\today}

\section{Introduction}
\label{Introduction}

In scalar-tensor theories  the gravitational interaction is mediated both by a metric tensor and a scalar field.
The interest in this type of theories of gravity is connected with the expectation that the observed late-time
accelerated expansion of the Universe may be understood without a dark-energy (DE) component \cite{carroll,nojiri1,gannouji1}.
Instead, it is the modified (compared with Einstein's theory) geometrical sector which is supposed to provide the desired dynamics\cite{copeland,torres,lobo,caldkam}.
This may be seen as a geometrization of DE.
Different aspects of scalar-tensor theories in general or subclasses of them have been investigated in
\cite{catena,farao,CAPOZ,dolgov,chiba,duma,sofarao,soti,brookfield,abean,scalten,clifton,chibayam,joyce}.

Scalar-tensor theories are formulated either in the Einstein frame or in the Jordan frame. Both frames are related by a conformal transformation.
While matter and scalar field energies are separately conserved in the Jordan frame, the dynamics of both components is coupled in the Einstein frame for any equation of state (EoS) different from that of radiation.
Because of the complex structure of scalar-tensor theories, simple solutions are difficult to obtain, even if
the symmetries of the cosmological principle are imposed. Hence, in practice, the background expansion rate is usually obtained via numerical integration of the equations of motion. In general, the scalar-tensor-theory based cosmological dynamics may substantially differ from standard cosmology.
Our focus here is on the simplest possible extension of the standard $\Lambda$CDM model that scalar-tensor theory can provide. In this minimalist approach we remain in the vicinity of the standard model at the present epoch and we aim to quantify the differences between scalar-tensor theory and general relativity (GR) by establishing a structure in which the scalar field $\Phi$
explicitly enters an analytic solution  of the dynamics  such that for $\Phi = 1$ the
standard $\Lambda$CDM limit is recovered.
To this purpose we construct a simple model which is analytically solved in the Einstein frame. With the help of a conformal transformation we then demonstrate how the field $\Phi$, which is given as a certain power of the scale factor, enters the (Jordan-frame) Hubble rate.
Here we rely on an effective GR description of the Jordan-frame dynamics to determine the geometric equivalent of DE.

In more detail, our starting point is a simple, analytically tractable expression for the coupling between nonrelativistic matter and the (Einstein frame) scalar field which modifies the standard decay of the matter energy density with the third power of the cosmic scale factor.
This interaction-triggered deviation of the standard decay in the Einstein frame is modeled by a power-law
behavior in terms of the Einstein-frame scale factor.
Under this condition and if additionally an effective energy density and an effective pressure, linked by a constant ``bare" EoS parameter, are assumed,
an explicit solution of the Einstein-frame dynamics is obtained with the mentioned power as an
additional parameter.
A straightforward conformal transformation then allows us to obtain the Hubble rate and the deceleration parameter in terms of this parameter in the Jordan frame as well.
For the value zero of such new parameter, corresponding to $\Phi = 1$, both frames become indistinguishable and reproduce the dynamics of the standard
$\Lambda$CDM model.
Otherwise one has a variable $\Phi$ and the dynamics in both frames becomes different, deviating from that
of the standard model.
The analytic expression for the Hubble rate which explicitly clarifies the impact of the scalar field on the cosmological dynamics is the main achievement of this paper.
We shall confront the deviations from the standard model with data from supernovae of type Ia (SNIa), the differential age of old galaxies that have evolved passively (using $H(z)$, where $H$ is the Hubble rate and $z$ is the redshift parameter) and baryon acoustic oscillations (BAO).

 The structure of the paper is as follows.
 In Sec.~\ref{Basic dynamics} we recall basic general relations for scalar-tensor theories and specify them to
 the homogeneous and isotropic case.
 In Sec.~\ref{two-component} we set up an effective two-component description in the Einstein frame,
 introduce our interaction model and find the Einstein-frame Hubble rate.
 The transformation to the Jordan frame is performed in Sec.~\ref{Jordan} where we also
 discuss the implications of a mapping of the scalar-field degrees of freedom on the effective fluid dynamics
 in a GR context.
 Section~\ref{observations} is devoted to a Bayesian statistical analysis on the basis of
 observational data of SNIa, $H(z)$ and BAO.
 Finally, in Sec.~\ref{conclusions} we summarize our results.



\section{Basics of scalar-tensor theories}
\label{Basic dynamics}

Scalar-tensor theories are based on the (Jordan-frame) action (see, e.g., \cite{abean,clifton,chibayam})
\begin{equation}
\label{action1}
   S(g_{\mu\nu},\Phi)= \frac{1}{2\kappa^{2}}\int d^{4}x\sqrt{-g}\left[\Phi R - \frac{\omega(\Phi)}{\Phi}\left(\nabla \Phi\right)^{2} - U(\Phi)\right] + S_{m}\left(g_{\mu\nu}\right)
\end{equation}
with a minimally coupled matter part
\begin{equation}\label{}
S_{m} = \int d^{4}x\sqrt{-g}L_{m}\left(g_{\mu\nu}\right)\,,
\end{equation}
where $\kappa^{2} = 8\pi G$ and  $L_{m}$ denotes the matter Lagrangian.
The dynamical field equations are
\begin{eqnarray}
\Phi\left(R_{\mu\nu}- \frac{1}{2}g_{\mu\nu}R\right) &=& \kappa^{2}T_{\mu\nu}
\nonumber\\
&& + \frac{\omega(\Phi)}{\Phi}
\left(\partial_{\mu}\Phi\partial_{\nu}\Phi
- \frac{1}{2}g_{\mu\nu}\left(\nabla\Phi\right)^{2}\right)
+ \nabla_{\mu}\nabla_{\nu}\Phi- g_{\mu\nu}\Box\Phi
- \frac{1}{2}g_{\mu\nu}U,
\end{eqnarray}
where the energy-momentum tensor of the matter is obtained as usual via
\begin{equation}\label{}
T_{\mu\nu} = - \frac{2}{\sqrt{-g}}\frac{\delta S_{m}}{\delta g^{\mu\nu}}.
\end{equation}
The dynamics of the field $\Phi$ is dictated by
\begin{equation}\label{}
\Box \Phi = \frac{1}{2\omega(\Phi) + 3}\left(\kappa^{2}T - \frac{d\omega(\Phi)}{d\Phi}\left(\nabla\Phi\right)^{2} + \Phi\frac{dU}{d\Phi} - 2U\right),
\end{equation}
which implies a coupling to the trace $T \equiv Tr (T_{\mu\nu})$ of the matter energy-momentum tensor.

Adopting the conformal transformation
\begin{equation}
g_{\mu\nu} = \frac{1}{\Phi}\tilde{g}_{\mu\nu} = e^{2\,b(\varphi)}\,\tilde{g}_{\mu\nu}
 \ \label{gtil}
\end{equation}
of the metric tensor $g_{\mu\nu} $ as well as a redefinition of the potential term
\begin{equation}\label{VU}
V(\varphi) = \frac{U(\Phi)}{2\kappa^{2}\Phi^{2}}
\end{equation}
and
\begin{equation}\label{Phivarphi}
\frac{1}{4\Phi^{2}}\left(\frac{d\Phi}{d\varphi}\right)^{2} = \left(\frac{d b}{d\varphi}\right)^{2}
= \frac{\kappa^{2}}{4\omega(\Phi)+6},
\end{equation}
one obtains the Einstein-frame action
\begin{equation}\label{}
S\left(\tilde{g}_{\mu\nu},\varphi\right) =
 \int d^{4}x\sqrt{-\tilde{g}}\left[\frac{1}{2\kappa^{2}}\,\tilde{R} - \frac{1}{2}\left(\tilde{\nabla} \varphi\right)^{2} - \tilde{V}\left(\varphi\right)
\right] + \int d^{4}x\sqrt{-\tilde{g}}\tilde{L}_{m}\left(e^{2b(\varphi)}\tilde{g}_{\mu\nu}\right),
 \ \label{n}
\end{equation}
in which the matter part is non-minimally coupled to the gravitational sector.
Throughout the paper, quantities without a tilde refer to the Jordan frame, quantities with a tilde have their meaning in the Einstein frame.

Restricting ourselves to spatially flat homogeneous and isotropic cosmological models with a Robertson-Walker metric and assuming a perfect-fluid structure for the energy-momentum tensor of the matter with energy density $\rho_{m}$, pressure $p_{m}$ and four-velocity $u^{\mu}$,
\begin{equation}\label{}
T_{\mu\nu} = \rho_{m}u_{\mu}u_{\nu} + p_{m}h_{\mu\nu},\quad h_{\mu\nu} = g_{\mu\nu} + u_{\mu}u_{\nu},
\end{equation}
the relevant Jordan-frame equations are
\begin{equation}\label{H2J}
H^{2} = \frac{\kappa^{2}}{3}\frac{\rho_{m}}{\Phi} + \frac{1}{3\Phi}
\left[\frac{1}{2}\frac{\omega(\Phi)}{\Phi}
  \left(\frac{\partial\Phi}{\partial t}\right)^{2}
  -3H\frac{\partial\Phi}{\partial t}
  + \frac{1}{2}U\right],
\end{equation}
where $H = \frac{1}{a}\frac{da}{dt}$ is the Hubble rate of the Jordan frame and
$a$ is the Jordan-frame scale factor,
\begin{equation}\label{}
\frac{d H}{d t} = - \frac{\kappa^{2}}{2} \frac{\rho_{m}+p_{m}}{\Phi}
- \frac{1}{2\Phi}\left[\frac{\omega(\Phi)}{\Phi}
  \left(\frac{\partial\Phi}{\partial t}\right)^{2}
  -H\frac{\partial\Phi}{\partial t}
  + \frac{d^{2}\Phi}{dt^{2}}\right],
\end{equation}
\begin{equation}\label{equPhigen}
\frac{d^{2}\Phi}{dt^{2}} + 3 H\frac{d\Phi}{dt} = \frac{1}{2\omega + 3}\left(\kappa^{2}\left(\rho_{m} - 3p_{m}\right)- \frac{d\omega}{d\Phi}\left(\frac{d\Phi}{dt}\right)^{2} - \Phi\frac{dU}{d\Phi} + 2U\right),
\end{equation}
as well as the matter conservation
\begin{equation}
\frac{d\rho_{m}}{dt} + 3H
\left(\rho_{m} + p_{m}\right) = 0.
\label{drtm}
\end{equation}
The corresponding relations for the Einstein frame are
\begin{equation}
\tilde{H}^{2} = \frac{\kappa^{2}}{3}\left[\tilde{\rho}_{m} + \frac{1}{2}\left(\frac{d\varphi}{d\tilde{t}}\right)^{2} + \tilde{V}\right],
\label{tH2}
\end{equation}
where   $\tilde{H} = \frac{1}{\tilde{a}}\frac{d\tilde{a}}{d\tilde{t}}$ is the Einstein-frame Hubble rate and $\tilde{a}$ is the Einstein-frame scale factor, as well as
\begin{equation}
\frac{d\tilde{H}}{d\tilde{t}} = - \frac{\kappa^{2}}{2}\left[\tilde{\rho}_{m} + \tilde{p}_{m} + \left(\frac{d\varphi}{d\tilde{t}}\right)^{2}
\right],
\label{dtH}
\end{equation}
\begin{equation}
\frac{d\tilde{\rho}_{m}}{d\tilde{t}} + 3\tilde{H} \left(\tilde{\rho}_{m} + \tilde{p}_{m}\right) =
- \frac{d\varphi}{d\tilde{t}}\frac{db}{d\varphi}
\left(-\tilde{\rho}_{m} +
3\tilde{p}_{m}\right)
\label{drmn}
\end{equation}
and
\begin{equation}
 \frac{d^{2}\varphi}{d\tilde{t}^{2}} + 3\tilde{H}\frac{d\varphi}{d\tilde{t}} + \tilde{V}_{,\varphi} =
\frac{db}{d\varphi}
\left(-\tilde{\rho}_{m} + 3\tilde{p}_{m}\right),
\label{ddvphi}
\end{equation}
where we have introduced the notation $\tilde{V}_{,\varphi} = \frac{\partial \tilde{V}}{\partial{\varphi}}$.

Different from the Jordan frame, the matter component is not separately conserved here.
The time coordinates and the scale factors of the FLRW metrics of both frames
are related by
\begin{equation}
dt =
e^{b} d \tilde{t}\ \qquad \mathrm{and} \qquad a =
e^{b}\tilde{a} \ ,\label{ttilb}
\end{equation}
respectively.
The matter pressure and the matter energy density transform as
\begin{equation}
p_{m} = e^{-4 b}\,\tilde{p}_{m} \qquad \mathrm{and} \qquad
\rho_{m} = e^{-4 b}\,\tilde{\rho}_{m}\,,
\label{mtilb}
\end{equation}
respectively.
This means $\frac{p_{m}}{\rho_{m}} = \frac{\tilde{p}_{m} }{\tilde{\rho}_{m}}$, i.e., the
EoS parameter of the matter remains invariant.

\section{Einstein-frame description}
\label{two-component}

\subsection{General relations}

The Einstein-frame equations (\ref{tH2}) -- (\ref{ddvphi}) are of an effective two-component structure in which matter interacts with a scalar field.
We associate an effective energy density $\tilde{\rho}_{\varphi}$ and an effective pressure $\tilde{p}_{\varphi}$ to the scalar field by
\begin{equation}
\tilde{\rho}_{\varphi} = \frac{1}{2}\left(\frac{d\varphi}{d\tilde{t}}\right)^{2}+ \tilde{V}\qquad \mathrm{and} \qquad \tilde{p}_{\varphi} =
\frac{1}{2}\left(\frac{d\varphi}{d\tilde{t}}\right)^{2} - \tilde{V},
\label{rhophi}
\end{equation}
respectively.
Equations (\ref{drmn}) and (\ref{ddvphi}) can then be written as
\begin{equation}
\frac{d\tilde{\rho}_{m}}{d\tilde{t}}  + 3\tilde{H} \left(1 + \tilde{w}_{m}\right)\tilde{\rho}_{m} = Q \equiv
\frac{d\varphi}{d\tilde{t}}\frac{d b}{d\varphi}
\left(1- 3\tilde{w}_{m}\right)\tilde{\rho}_{m}
 \ \label{drm+b}
\end{equation}
and
\begin{equation}
\frac{d\tilde{\rho}_{\varphi}}{d\tilde{t}}  + 3\tilde{H} \left(1 +
\tilde{w}\right)\tilde{\rho}_{\varphi} = - Q
= - \frac{d\varphi}{d\tilde{t}}\frac{d b}{d\varphi}
\left(1- 3\tilde{w}_{m}\right)\tilde{\rho}_{m},
\label{drp+b}
\end{equation}
respectively, where $\tilde{w}_{m} =
\frac{\tilde{p}_{m}}{\tilde{\rho}_{m}}$ is the matter EoS parameter
and $\tilde{w} = \frac{\tilde{p}_{\varphi}}{\tilde{\rho}_{\varphi}}$ is the Einstein-frame EoS parameter for the scalar field. Notice the difference in notation between the EoS parameter $w$ and the coupling $\omega$ in the action of the scalar-tensor theory.
The interaction vanishes for the special case $\tilde{w}_{m} = 1/3$.
The total energy density $\tilde{\rho}$ and the total pressure $\tilde{p}$ are
\begin{equation}
\tilde{\rho} = \tilde{\rho}_{m} + \tilde{\rho}_{\varphi},\qquad  \tilde{p} = \tilde{p}_{m} + \tilde{p}_{\varphi},
\label{rho}
\end{equation}
for which the conservation equation
\begin{equation}
\frac{d\tilde{\rho}}{d\tilde{t}} + 3\tilde{H} \left(\tilde{\rho} + \tilde{p}\right) = 0
 \,\label{rho}
\end{equation}
holds.

\subsection{Modeling the interaction}

In the special case $\tilde{p}_{m}=0$
the solution of (\ref{drm+b}) can be written as
\begin{equation}
\tilde{\rho}_{m} = \tilde{\rho}_{m0}\tilde{a}^{-3}f(\tilde{a}) \quad\Rightarrow \quad
\frac{d\tilde{\rho}_{m}}{d\tilde{t}}  + 3\tilde{H} \tilde{\rho}_{m} = \frac{\tilde{\rho}_{m}}{f}
\frac{d f}{d\tilde{t}}\,,
 \ \label{rhof}
\end{equation}
where the function $f$ encodes the effects of an interaction between matter and scalar field.
Comparing (\ref{drm+b}) and (\ref{rhof}), it follows that
\begin{equation}\label{dfb}
\frac{1}{f}
\frac{d f}{d\tilde{t}} = \frac{d\varphi}{d\tilde{t}}\frac{d b}{d\varphi}
\quad
\Rightarrow\quad f = e^{b(\varphi)}.
\end{equation}
The absence of an interaction means a constant $f$, equivalent to $\varphi = \varphi_{0} = $ constant.
Writing the time derivative of $f\left(\tilde{a}\right)$ as
\begin{equation}
\label{f}
\frac{d f}{d\tilde{t}} = \frac{d f}{d \tilde{a}}\frac{d \tilde{a}}{d\tilde{t}}
= \frac{d f}{d \tilde{a}}\tilde{a}\tilde{H}\,,
\end{equation}
Eq. (\ref{drp+b})  becomes
\begin{equation}\label{}
\frac{d\tilde{\rho}_{\varphi}}{d \ln\tilde{a}} + 3\left(1+\tilde{w}\right)\tilde{\rho}_{\varphi}
= - \tilde\rho_{m_0}\tilde{a}^{-3}\frac{d f}{d \ln\tilde{a}}\ .
\end{equation}
For a constant EoS parameter $\tilde{w}$ the solution of this equation is
\begin{equation}
\tilde{\rho}_{\varphi} = \tilde{\rho}_{\varphi 0}\tilde{a}^{-3\left(1+\tilde{w}\right)} - \tilde\rho_{m_0}\,\tilde{a}^{-3\left(1+\tilde{w}\right)}\, \int_{\tilde{a}_{0}}^{\tilde{a}} d\tilde{a}
\frac{d f}{d \tilde{a}}\,\tilde{a}^{3\tilde{w}}\ .
\label{rx3}
\end{equation}
For a given interaction $f(\tilde{a})$, the Hubble rate (\ref{tH2}) is then determined by the sum of  $\tilde{\rho}_{m}$ from (\ref{rhof}) with (\ref{f}) and $\tilde{\rho}_{\varphi}$ from (\ref{rx3}).

To construct an explicitly solvable model we shall assume a power-law behavior of the interaction function $f\left(\tilde{a}\right)$,
\begin{equation}\label{pansatz}
f\left(\tilde{a}\right) = \tilde{a}^{3m}.
\end{equation}
Clearly, for $m=0$ the function $f$ becomes constant and the interaction is absent.

With (\ref{pansatz}) the explicit solution of (\ref{rx3}) then is
\begin{equation}
\tilde{\rho}_{\varphi}= \left(\tilde{\rho}_{\varphi 0} + \frac{m}{\tilde{w} + m}\tilde\rho_{m_0} \right)\tilde{a}^{-3\left(1+\tilde{w}\right)} - \frac{m}{\tilde{w} + m}\tilde\rho_{m_0}\tilde{a}^{-3\left(1-m\right)}
.
\label{tilrhophi}
\end{equation}
Via
\begin{equation}\label{dvarphi3m}
\frac{d\varphi}{d\tilde{t}}\frac{db}{d\varphi} = 3m\tilde{H}, \qquad \frac{d\varphi}{d\ln\tilde{a}}\frac{db}{d\varphi} = 3m,
\end{equation}
the interaction term $Q$ becomes
\begin{equation}\label{Qm}
Q = 3m\tilde{H}\tilde{\rho}_{m}.
\end{equation}

\subsection{Effective EoS and Hubble expansion rate for a linear dependence $b(\varphi)=K\varphi$}

In order to obtain analytical expressions for the dynamics we consider the simple case of a linear dependence
\begin{equation}\label{blin}
b=b(\varphi) = K\varphi, \qquad K = \sqrt{\frac{\kappa^{2}}{4\omega +6}},
\end{equation}
where the expression for $K$ follows from  (\ref{Phivarphi}).
This implies also the relations
\begin{equation}\label{varphitila}
\varphi=\frac{3m}{K}\ln\tilde{a},\quad b=3m \ln\tilde{a}, \quad e^{b} = \tilde{a}^{3m},\quad \frac{db}{d\tilde{t}} = 3m\tilde{H},
\end{equation}
as well as
\begin{equation}\label{Phivarphi+}
\Phi = e^{-2K\varphi} = e^{-2b} =\tilde{a}^{-6m}.
\end{equation}
The power $m$ is a direct measure of the interaction strength.
The interaction-free case $m=0$ corresponds to $\varphi = \varphi_{0}= $ constant, i.e.,
 $\tilde{V} = \tilde{\rho}_{\varphi}$ and, consequently, to $\tilde{w} = -1$, equivalent to the $\Lambda$CDM model.
For $m>0$ one has $Q>0$ while for $m<0$ the opposite, $Q<0$, is valid.
The first equation in (\ref{Phivarphi+}) relates the scalars $\Phi$ and $\varphi$ without specifying a potential $V(\varphi)$.

For the special case of an exponential
potential,
\begin{equation}\label{}
V(\varphi) = V_{0}e^{-\lambda\varphi},
\end{equation}
one has, from (\ref{Phivarphi}),
\begin{equation}\label{}
\varphi = - \frac{1}{2K}\ln \Phi \quad \Rightarrow\quad V= V_{0}\Phi^{\frac{\lambda}{2K}}
\end{equation}
and relation (\ref{VU}) provides us with the power-law potential
\begin{equation}\label{}
U(\Phi) = U_{0}\Phi^{2+\frac{\lambda}{2K}}, \qquad U_{0} =2\kappa^{2}V_{0}.
\end{equation}

With (\ref{dvarphi3m}) the balance equation (\ref{drp+b}) for $\tilde{\rho}_{\varphi}$ can be written
\begin{equation}
\frac{d\tilde{\rho}_{\varphi}}{d\tilde{t}}  + 3\tilde{H} \left(1 +
\tilde{w}^{eff}\right)\tilde{\rho}_{\varphi} = 0,
\label{drpeff}
\end{equation}
with an effective EoS parameter
\begin{equation}\label{tilweff}
\tilde{w}^{eff} = \tilde{w} + m\frac{\tilde{\rho}_{m}}{\tilde{\rho}_{\varphi}},
\end{equation}
corresponding to an effective pressure $\tilde{p}_{\varphi}^{eff} = \tilde{w}^{eff}\tilde{\rho}_{\varphi}$.
The interaction modifies the ``bare" EoS parameter $\tilde{w}$. The modification is linear in the interaction parameter $m$.
Likewise, the matter balance (\ref{drm+b}) takes the form of a conservation equation
\begin{equation}
\frac{d\tilde{\rho}_{m}}{d\tilde{t}}  + 3\tilde{H} \left(1 - m\right)\tilde{\rho}_{m} = 0
\qquad \Rightarrow \qquad \tilde{\rho}_{m} = \tilde{\rho}_{m0}\tilde{a}^{-3\left(1 - m\right)}.
\label{}
\end{equation}
Since from (\ref{Phivarphi+}) one has  $\tilde{a} = e^{\frac{K}{3m}\varphi}$,
it follows that the matter energy density can be written in terms of the scalar field as
\begin{equation}\label{rhomphib}
\tilde{\rho}_{m} = \tilde{\rho}_{m0}\tilde{a}^{-3\left(1-m\right)}
= \tilde{\rho}_{m0}e^{-\frac{1-m}{m}K\varphi}.
\end{equation}
This exponential structure allows us to write
\begin{equation}\label{}
\tilde{\rho}_{m} = - \frac{m}{K\left(1-m\right)}\tilde{\rho}_{m,\varphi},
\end{equation}
which is of interest in defining an effective potential.
Then, Eq.~(\ref{drp+b}) (for $\tilde{p}_{m} = 0$) is equivalent to
\begin{equation}
 \frac{d^{2}\varphi}{d\tilde{t}^{2}} + 3\tilde{H}\frac{d\varphi}{d\tilde{t}} + \tilde{V}^{eff}_{,\varphi}
 = 0,
\label{ddphieff}
\end{equation}
where $\tilde{V}^{eff}$ now includes the interaction,
\begin{equation}\label{}
\tilde{V}^{eff} = \tilde{V} + \tilde{V}_{int} \equiv \tilde{V} - \frac{m}{\left(1-m\right)}\tilde{\rho}_{m}.
\end{equation}
Because of the representation in (\ref{rhomphib}), the interaction potential is of the exponential type.
One may introduce effective quantities
\begin{equation}
\tilde{\rho}_{\varphi}^{eff} = \frac{1}{2}\left(\frac{d\varphi}{d\tilde{t}}\right)^{2}+ \tilde{V}^{eff}\qquad \mathrm{and} \qquad \tilde{p}_{\varphi}^{eff} =
\frac{1}{2}\left(\frac{d\varphi}{d\tilde{t}}\right)^{2} - \tilde{V}^{eff},
\label{rhophieff}
\end{equation}
for which, from (\ref{ddphieff}),
\begin{equation}
\frac{d\tilde{\rho}_{\varphi}^{eff}}{d\tilde{t}}  + 3\tilde{H} \left(1 +
\tilde{W}^{eff}\right)\tilde{\rho}_{\varphi}^{eff} = 0
\label{drpeff2}
\end{equation}
is valid with
\begin{equation}\label{}
\tilde{W}^{eff} = \frac{\tilde{p}_{\varphi}^{eff}}{\tilde{\rho}_{\varphi}^{eff}}
= \frac{\tilde{p}_{\varphi} + \frac{m}{1-m}\tilde{\rho}_{m}}{\tilde{\rho}_{\varphi} - \frac{m}{1-m}\tilde{\rho}_{m}} = \frac{\tilde{w} + \frac{m}{1-m}r}{1 -\frac{m}{1-m}r}.
\end{equation}
For $\tilde{w} = - 1$ we have $\tilde{W}^{eff} = - 1$ as well.

Together with (\ref{rhomphib}) the total energy becomes
\begin{equation}\label{}
\tilde{\rho} = \tilde{\rho}_{m} + \tilde{\rho}_{\varphi} =
\frac{1}{1 + m/\tilde{w}}\tilde{\rho}_{m0}\tilde{a}^{-3\left(1-m\right)}
+ \left(\tilde{\rho}_{\varphi 0} + \frac{m}{\tilde{w} + m}\tilde\rho_{m_0} \right)\tilde{a}^{-3\left(1+\tilde{w}\right)}.
\end{equation}
For the actual value we have $\tilde{\rho}_{0} = \tilde{\rho}_{m0} + \tilde{\rho}_{\varphi 0}$.
Introducing the fractional quantities
\begin{equation}\label{}
\tilde{\Omega}_{m0} = \frac{8\pi G\tilde{\rho}_{m0}}{3\tilde{H}_{0}^{2}} \ , \quad
\tilde{\Omega}_{\varphi 0} = \frac{8\pi G\tilde{\rho}_{\varphi 0}}{3\tilde{H}_{0}^{2}} = 1 - \tilde{\Omega}_{m0},
\end{equation}
the square of the Hubble rate is
\begin{equation}\label{tilh2}
\frac{\tilde{H}^{2}}{\tilde{H}^{2}_{0}} = \frac{1}{1 + m/\tilde{w}}\left\{\tilde{\Omega}_{m0}\tilde{a}^{-3\left(1-m\right)}
+ \left[1 + \frac{m}{\tilde{w}} - \tilde{\Omega}_{m0}\right]\tilde{a}^{-3\left(1+\tilde{w}\right)}\right\}.
\end{equation}
The non-interacting case $m=0$ corresponds to a $w$CDM model. As already mentioned, since then $\varphi = $ constant, the only possibility here is $\tilde{w} = -1$, i.e., the $\Lambda$CDM model.

In the following we shall focus on the case $\tilde{w} = -1$ but admitting $m$ to be non vanishing. Under this condition $\tilde{H}^{2}$ consists of a constant part like the $\Lambda$CDM model but modified by the presence of $m$ and
a matter part in which the parameter $m$ modifies the conventional $\tilde{a}^{-3}$ behavior:
\begin{equation}\label{tilh2+}
\frac{\tilde{H}^{2}}{\tilde{H}^{2}_{0}} = e^{K\varphi}\frac{\tilde{\Omega}_{m0}\tilde{a}^{-3}}{1 - m}
+ 1  - \frac{\tilde{\Omega}_{m0}}{1 - m}\ .
\end{equation}
These small modifications make the Einstein-frame dynamics different from that of the $\Lambda$CDM model.
(Alternatively, this solution may be interpreted in a purely GR context with the $\varphi$ component belonging to the matter part of the field equations, interacting with nonrelativistic matter.)

For the deceleration parameter
\begin{equation}\label{}
\tilde{q} = - 1 - \frac{\tilde{a}}{\tilde{H}}\frac{d\ \tilde{H}}{d\ \tilde{a}}\
\end{equation}
we find
\begin{equation}\label{tilq}
\tilde{q}
= \frac{1}{2}\frac{\left(1-3m\right)\tilde{\Omega}_{m0}\tilde{a}^{-3\left(1- m\right)}
- 2\left[1 - m  - \tilde{\Omega}_{m0}\right]
}{\tilde{\Omega}_{m0}\tilde{a}^{-3\left(1- m\right)}
+ \left[1 - m - \tilde{\Omega}_{m0}\right]}\,.
\end{equation}
For $m = 0$  its present value consistently reduces
 to $\tilde{q}_{0} = -1 + \frac{3}{2}\tilde{\Omega}_{m0}$.

\subsection{Consistency check}

 Now, a consistency check can be performed as follows. With (\ref{varphitila}) we have
\begin{equation}\label{dvarphiK}
\left(\frac{d\varphi}{d\tilde{t}}\right)^{2} = \frac{9m^{2}}{K^{2}}\tilde{H}^{2}.
\end{equation}
Taking into account $\tilde{H}^{2} = \frac{\kappa^{2}}{3}\left(\tilde{\rho}_{m} + \tilde{\rho}_{\varphi}\right)$ results in
\begin{equation}\label{}
\tilde{w} = \frac{\frac{1}{2}\left(\frac{d\varphi}{d\tilde{t}}\right)^{2} - V}{\frac{1}{2}\left(\frac{d\varphi}{d\tilde{t}}\right)^{2} + V}
=\frac{3m^{2}\left(2\omega +3\right)\left(\tilde{\rho}_{m} + \tilde{\rho}_{\varphi}\right) - V}
{ 3m^{2}\left(2\omega +3\right)\left(\tilde{\rho}_{m} + \tilde{\rho}_{\varphi}\right) + V},
\end{equation}
equivalent to
\begin{equation}\label{tw}
\tilde{w} = - 1 + \frac{6m^{2}\left(2\omega +3\right)\frac{\left(\tilde{\rho}_{m} + \tilde{\rho}_{\varphi}\right)}{V}}{1+3m^{2}\left(2\omega +3\right)\frac{\left(\tilde{\rho}_{m} + \tilde{\rho}_{\varphi}\right)}{V}}
= - 1 + \mathcal{O}(m^{2}).
\end{equation}
In general, our initial assumption of a constant $\tilde{w}$ does not seem to be compatible with the dynamics of $\varphi$ in (\ref{dvarphiK}).
However, the corrections to the constant value $\tilde{w} = -1$ are quadratic in the interaction parameter $m$,
there does not appear a term linear in $m$ in (\ref{tw}). Our approach will therefore be consistent, if we restrict ourselves to $\tilde{w} = -1$
and to modifications of the effective equation of state (\ref{tilweff}) which are linear in $m$.
Since $m$ quantifies deviations from the $\Lambda$CDM model, one expects $m$ to be small.

\section{Jordan-frame description}
\label{Jordan}

With $H = \frac{1}{a}\frac{da}{dt}$ the transformations (\ref{ttilb}) allow us to establish the relation between
the Hubble rates of both frames:
\begin{equation}\label{}
H= e^{-b}\left(1+3m\right)\tilde{H}.
\end{equation}
With
\begin{equation}\label{}
f = e^{b} = \frac{a}{\tilde{a}}
\end{equation}
and (\ref{pansatz}) we have
\begin{equation}\label{ata}
a = \tilde{a}^{3m+1}\quad \Rightarrow\quad \tilde{a} = a^{\frac{1}{3m+1}} \quad \Rightarrow\quad
\frac{\tilde{a}}{a} = a^{-\frac{3m}{3m+1}}\,,
\end{equation}
i.e., the power $m$ quantifies the difference of the scale factors in both frames.
For $m=0$ one has $\tilde{a} = a$ and the dynamics in both frames reduces to that of the $\Lambda$CDM model.
Combining (\ref{varphitila}) and (\ref{ata}) yields $b$ in terms of the Jordan-frame scale factor $a$,
\begin{equation}\label{}
  b= \frac{3m}{1+3m}\ln a , \qquad \frac{db}{dt} = \frac{3m}{1+3m}H.
\end{equation}
For the Hubble rate it follows that
\begin{equation}\label{HtH}
H = \frac{\tilde{a}}{a}\left(1+3m\right)\tilde{H}.
\end{equation}
The matter energy densities in both frames are related by
\begin{equation}\label{}
\rho_{m} = e^{-4b}\tilde{\rho}_{m} =\left(\frac{\tilde{a}}{a}\right)^{4} \tilde{\rho}_{m0}\tilde{a}^{-3(1-m)}
= a^{-\frac{12m}{1+3m}}\tilde{\rho}_{m0}\tilde{a}^{-3(1-m)} = \rho_{m0}a^{-3},
\end{equation}
where we have used $\tilde{\rho}_{m0} = \rho_{m0}$. As expected, we recover the characteristic
$a^{-3}$ behavior for separately conserved non-relativistic matter in the Jordan frame.

Using (\ref{ata}) and (\ref{tilh2}) in (\ref{HtH}) we obtain the Jordan-frame Hubble rate square
\begin{equation}\label{}
H^{2} = \left(1+3m\right)^{2} \tilde{H}_{0}^{2}
\left\{\frac{\tilde{\Omega}_{m0}}{1+m/\tilde{w}}a^{-\frac{3m+3}{3m+1}}
+ \left[1 - \frac{\tilde{\Omega}_{m0}}{1+m/\tilde{w}}a^{-\frac{6m + 3\left(1+\tilde{w}\right)}{3m+1}}\right]\right\}
\,.
\end{equation}
Obviously, $H^{2}_{0} = \left(1+3m\right)^{2} \tilde{H}_{0}^{2}$.
This implies
$\tilde{\Omega}_{m0} =  \left(1+3m\right)^{2}\Omega_{m0}$. Then, for $\tilde{w}=-1$,
\begin{equation}\label{H2}
\frac{H^{2}}{H_{0}^{2}} =
A\Omega_{m0}a^{-3+\frac{6m}{1+3m}}
+ \left[1 - A\Omega_{m0}\right]a^{-\frac{6m}{1+3m}},\qquad A\equiv \frac{\left(1 +3m \right)^{2}}{1 - m} ,
\end{equation}
or, since
\begin{equation}\label{}
a^{\frac{6m}{1+3m}} = e^{2b} = \Phi^{-1},
\end{equation}
the square of the Hubble rate can be written as
\begin{equation}\label{H2Phi}
\frac{H^{2}}{H_{0}^{2}} =
\frac{A\Omega_{m0}a^{-3}}{\Phi}
+ \left[1 - A\Omega_{m0}\right]\Phi .
\end{equation}
This expression is our main result.  It represents the (Jordan-frame) Hubble rate of our scalar-tensor-theory model.
From the structure of (\ref{H2Phi}) it is obvious,  how the scalar $\Phi$ modifies the cosmological dynamics compared
with the GR based standard model.
For $m=0$, equivalent to $\Phi = 1$, we recover the $\Lambda$CDM model and the Jordan-frame dynamics coincides with the dynamics in the Einstein frame.
For any $m\neq 0$, equivalent to $\Phi \neq 1$, the expression (\ref{H2Phi}) (or (\ref{H2})) represents a testable, alternative model with presumably small deviations from the $\Lambda$CDM model.
Notice that the modifications of the $\Lambda$CDM model are different from those in the Einstein frame.
In particular, there is no constant part even for $\tilde{w}=-1$.

It should be emphasized that the solution $\Phi = a^{-\frac{6m}{1+3m}}$ is a consequence of the solution of the macroscopic fluid dynamical equations for the energy densities under the assumption (\ref{pansatz}). It is not a solution of equation (\ref{equPhigen}), which would require an explicit expression for the potential $U$.
The unknown exact solution of the scalar field equation (\ref{equPhigen}) is replaced here by the effective solution $\Phi = a^{-\frac{6m}{1+3m}}$ which was obtained using the approximations (\ref{pansatz}) and (\ref{blin}) in the macroscopic fluid dynamics. Our procedure allows us to obtain an explicit expression for $\Phi$ and for the Hubble rate without solving the scalar field equation (\ref{equPhigen}). This can be seen as a major advantage of the approach.
By direct calculation one verifies that $\Phi$ obeys the simplified effective equation of motion
\begin{equation}\label{equeff}
\frac{d^{2}\Phi}{dt^{2}} + 3H\frac{d\Phi}{dt}
= - 9H_{0}^{2} \frac{m}{\left(1+3m\right)^{2}}\left[\left(1+m\right)A\Omega_{m0}a^{-3}
+ 2\left(1 - A\Omega_{m0}\right)\Phi^{2}\right]
\end{equation}
instead of (\ref{equPhigen}).
Since
\begin{equation}\label{}
a^{-3} = \Phi^{\frac{1+3m}{2m}},
\end{equation}
equation (\ref{equeff}) is of the form
\begin{equation}\label{eqPhieff}
\frac{d^{2}\Phi}{dt^{2}} + 3H\frac{d\Phi}{dt} + U^{eff}_{,\Phi} = 0,
\end{equation}
where
\begin{equation}\label{Ueff}
U^{eff}_{,\Phi} = 9H_{0}^{2} \frac{m}{\left(1+3m\right)^{2}}\left[\left(1+m\right)A\Omega_{m0}\Phi^{\frac{1+3m}{2m}}
+ 2\left(1 - A\Omega_{m0}\right)\Phi^{2}\right].
\end{equation}
This means, the dynamics of our $\Phi$ is that of a scalar field with potential $U^{eff}$. For $m=0$ the derivative $U^{eff}_{,\Phi}$ vanishes, corresponding to a constant effective potential and compatible with
$\Phi = 1$, thus recovering the $\Lambda$CDM model. It should be kept in mind, however, that the dynamics of $\Phi$
describes deviations from the $\Lambda$CDM model, not this model itself.

The total energy density corresponding to (\ref{H2Phi}) may be written as
\begin{equation}\label{}
\rho = A\frac{\rho_{m}}{\Phi} + \rho_{0}\left(1-A\Omega_{m0}\right)\Phi ,
\end{equation}
or, separating the matter part as in (\ref{H2J})
\begin{equation}\label{}
  \rho = \frac{\rho_{m}}{\Phi} + \left(A-1\right)\frac{\rho_{m}}{\Phi} + \rho_{0}\left[1 - A\Omega_{m0}\right]\Phi .
\end{equation}
The appearance of the scalar $\Phi$ in these expressions changes the relative contributions of matter and the dark-energy equivalent.
Our model encodes the deviations of the scalar-tensor description from the $\Lambda$CDM model entirely in the constant parameter $m$ which is supposed to be small. To be more definite we shall assume $|m|<\frac{1}{3}$. Under this condition $\Phi$ decays with $a$ for $m>0$ while it increases for $m<0$. It is only exactly at the present epoch $a=1$ that $\Phi (a=1)=1$.
For $m>0$ we have $\Phi > 1$ in the past ($a< 1$), for $m<0$, on the other hand, $\Phi$ increases from $\Phi < 1$ at $a< 1$ to the present $\Phi (a=1)=1$.
For $a\ll 1$ the energy density approaches
\begin{equation}\label{}
\rho \approx A \rho_{m0}a^{-3} a^{\frac{6m}{1+3m}} \qquad\qquad\qquad (a\ll 1).
\end{equation}
In the far-future limit $a\gg 1$ we have
\begin{equation}\label{rhofut}
\rho \approx \rho_{0}\left(1-A\Omega_{m0}\right)a^{-\frac{6m}{1+3m}}\qquad\qquad (a\gg 1).
\end{equation}
Depending on the sign of $m$ it may either decay ($m>0$) or increase ($m<0$).

The matter fraction becomes
\begin{equation}\label{}
\Omega_{m} = \frac{\rho_{m}}{\rho} = \frac{\Omega_{m0}a^{-3}}{A\Phi^{-1}\Omega_{m0}a^{-3} + \left[1 - A\Omega_{m0}\right]\Phi}.
\end{equation}
We may introduce an effective GR description by defining a component $\rho_{x} = \rho - \rho_{m}$,
\begin{equation}\label{rhox}
\frac{\rho_{x}}{\rho_{0}} = \left[1 - A\Omega_{m0}\right]a^{-\frac{6m}{1+3m}} + \Omega_{m0}a^{-3}
\left[A\,a^{\frac{6m}{1+3m}} - 1\right],
\end{equation}
equivalent to
\begin{equation}\label{}
\rho_{x} = \left(\frac{A}{\Phi}-1\right)\rho_{m} + \rho_{0}\left[1 - A\Omega_{m0}\right]\Phi
\end{equation}
with the fractional contribution
\begin{equation}\label{}
\Omega_{x} = \frac{\rho_{x}}{\rho} = \frac{\left(A\Phi^{-1} -1\right)\Omega_{m0}a^{-3} + \left[1 - A\Omega_{m0}\right]\Phi}
{A\Phi^{-1}\Omega_{m0}a^{-3} + \left[1 - A\Omega_{m0}\right]\Phi}.
\end{equation}
Then, at high redshift,
\begin{equation}\label{}
\Omega_{m} = \frac{\rho_{m}}{\rho} \approx \frac{1}{A}a^{-\frac{6m}{1+3m}},\qquad \Omega_{x} = \frac{\rho_{x}}{\rho}
\approx 1 - \frac{1}{A}a^{-\frac{6m}{1+3m}}
\qquad\qquad (a\ll 1).
\end{equation}
For $m>0$ the effective energy density $\rho_{x}$ becomes negative for small values of $a$.
While the combination $1 - A\Omega_{m0}$ remains always positive for small values of $m$ and $\Omega_{m0}$ of the order of 0.3, the combination
$A\,a^{\frac{6m}{1+3m}} - 1$ will change its sign with the increase of the scale factor. This sign change will occur at a value $a_{c}$, given by
\begin{equation}\label{}
A\,a_{c}^{\frac{6m}{1+3m}} - 1 = 0 \quad \Rightarrow \quad
a_{c} = \left[\frac{1-m}{\left(1+3m\right)^{2}}\right]^{\frac{1+3m}{6m}}.
\end{equation}
It is straightforward to  check that values $ \mathcal{O}\left|m\right| \gtrsim 0.1$ drastically change the dynamics compared with that of the standard model. Namely, for $m> 0$ the combination $A\,a^{\frac{6m}{1+3m}} - 1$ is negative for small values of $a$. With increasing $a$ one obtains a change to positive values at $a_{c} \approx 0.25$ for $m=0.1$, corresponding to a redshift $z_{c}\approx 3$ and at $a_{c} \approx 0.31$ (redshift $z_{c}\approx 2.2$) for $m=0.01$.
For $m< 0$ the term $A\,a^{\frac{6m}{1+3m}} - 1$ is positive for $a\ll 1$ but will change the sign as well. For $m=-0.1$ this happens at $a_{c} \approx 0.39$ (redshift $z_{c}\approx 1.6$), for $m=-0.01$ the corresponding values are $a_{c} \approx 0.08$ ($z_{c}\approx 11.6$).
The behavior of the fractional energy densities is shown in Figs.~1. The right panels use logarithmic units showing the asymptotic values of $\Omega_m$ and $\Omega_{x}$ at early times.
It should be emphasized again that the ``energy density" $\rho_{x}$ does not represent a material substratum, it is of geometric origin.
The fact that the quantity $\rho_{x}$ may become negative does not jeopardize the model.
What matters here is the Hubble rate which is always well behaved.

\begin{figure}
\begin{center}
\includegraphics[width=0.4\textwidth]{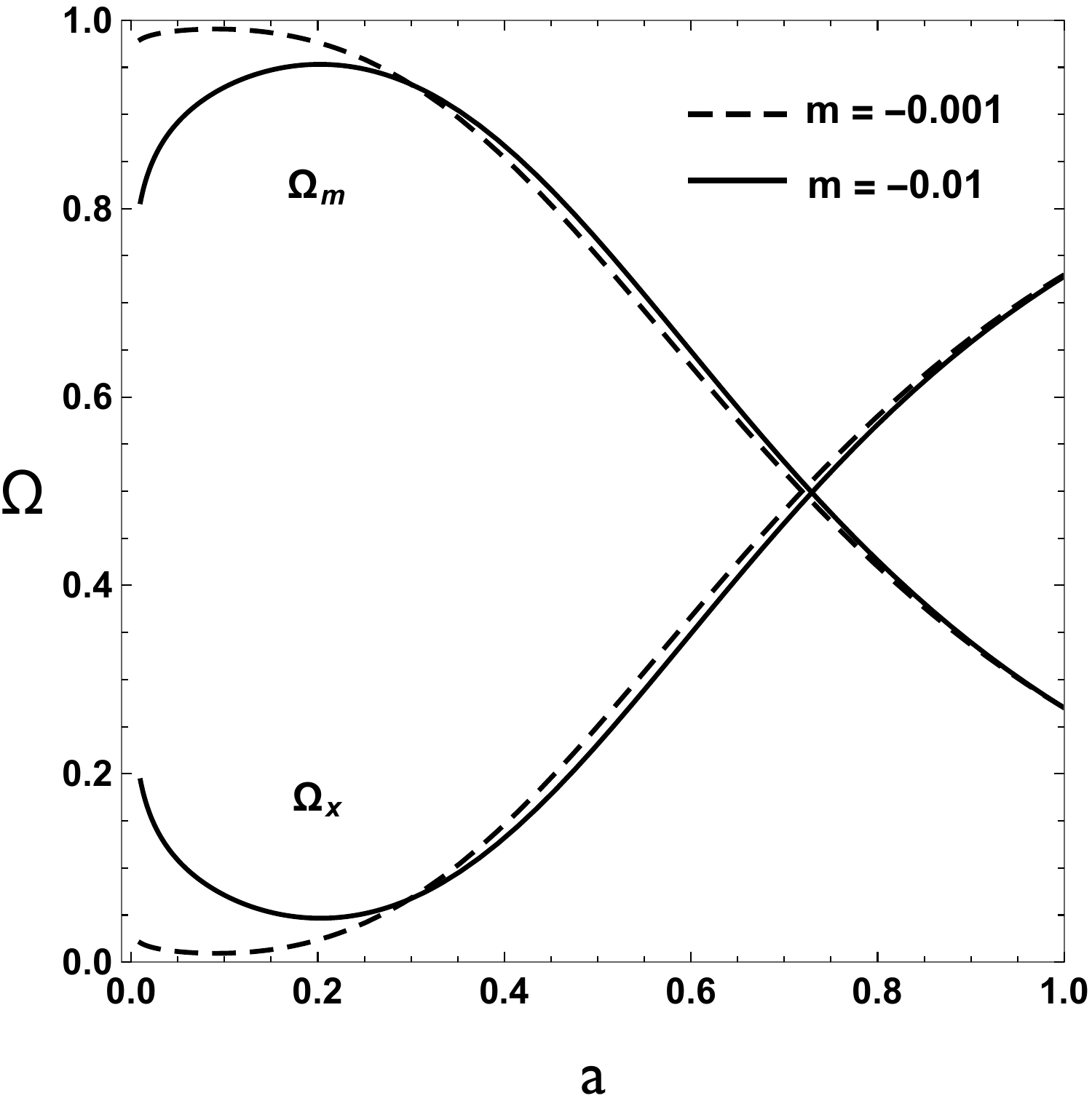}
\hspace{1cm}\includegraphics[width=0.4\textwidth]{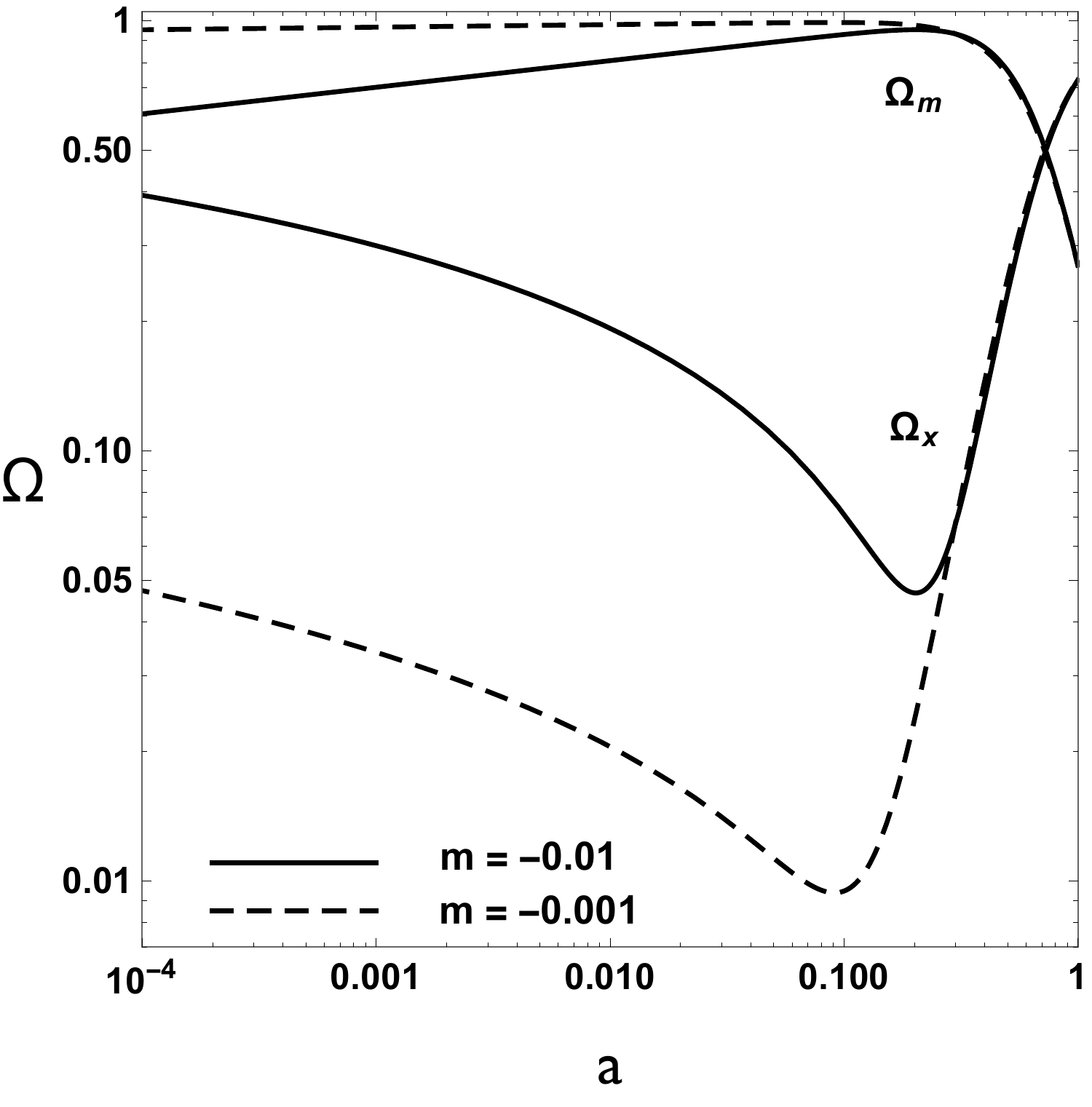}
\includegraphics[width=0.4\textwidth]{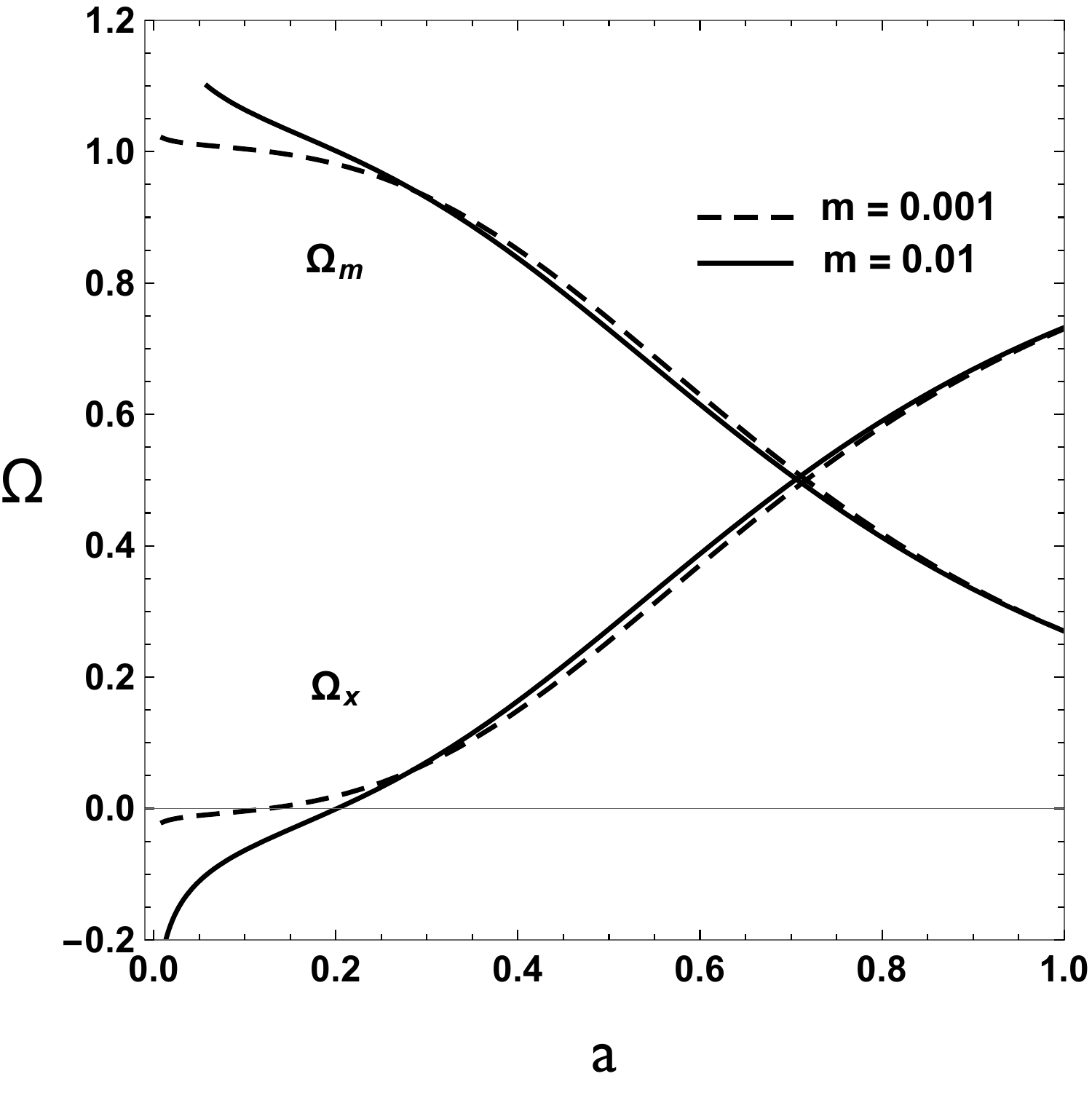}
\hspace{1cm}\includegraphics[width=0.4\textwidth]{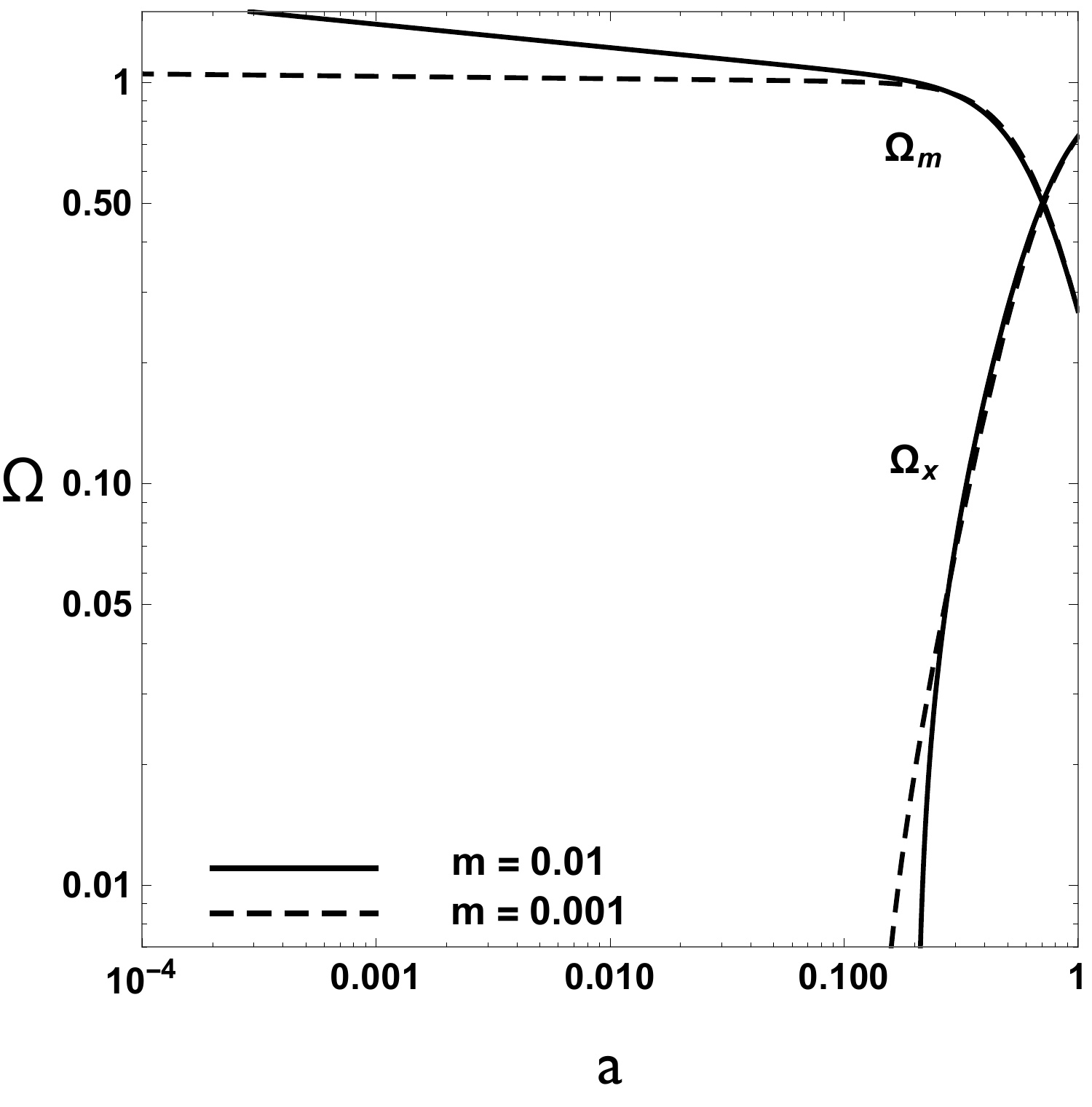}
\label{figOmxn}	
\caption{Matter fraction $\Omega_m$ of the total energy and geometric energy fraction $\Omega_x$ for negative (top panels) and positive (bottom panels) values of $m$. Logarithmic units are used in the right panels. This allows for a better visualization for the asymptotic values at early times.  }
\end{center}
\end{figure}



To complete the analogy, the component $\rho_{x}$ is supposed to obey the conservation equation
\begin{equation}\label{}
\frac{d \rho_{x}}{d t} + 3H\left(1 + w_{x}\right)\rho_{x} = 0\
\end{equation}
with an effective EoS parameter $w_{x}$.
With $
\frac{d \rho_{x}}{d t} = \frac{d \rho_{x}}{d a}\frac{d a}{d t} \equiv \rho_{x}^{\prime}aH$
we obtain
\begin{equation}\label{}
\rho_{x}^{\prime}a + 3\left(1 + w_{eff}\right)\rho_{x} = 0 \quad \Rightarrow\quad
1 + w_{x} = - \frac{1}{3}\frac{\rho_{x}^{\prime}a}{\rho_{x}}.
\end{equation}
A direct calculation yields
\begin{equation}\label{wtil-1}
w_{x} = - 1 + \frac{\frac{2m}{1+3m}\left[1 - A\Omega_{m0}\right]\Phi + \Omega_{m0}a^{-3}\left[\frac{1+m}{1+3m}A \Phi^{-1} - 1\right]}{\left[1 - A\Omega_{m0}\right]\Phi + \Omega_{m0}a^{-3}
\left[A \Phi^{-1} - 1\right]}.
\end{equation}
Equation (\ref{wtil-1}) establishes a relation between the constant, ``bare" EoS parameter $\tilde{w}=-1$ in the
Einstein frame and the generally time-dependent effective Jordan-frame EoS parameter $w_{x}$.
 For $m=0$ we recover $w_{x} = \tilde{w} = -1$.
At high redshift
 \begin{equation}\label{}
w_{x} \approx - 1 + \frac{\left[\frac{1+m}{1+3m}A \Phi^{-1} - 1\right]}{\left[A \Phi^{-1} - 1\right]}
\qquad\qquad (a\ll 1).
\end{equation}
 The value of $w_{x}$ may be close to zero at high redshift, i.e., mimicking dust,
but the effective energy density will be negative for $m>0$.
 Already a rather small value of $|m| \neq 0$ will substantially change the behavior of this dark-energy equivalent compared with the standard-model dark energy.
The present value of the effective EoS parameter is
\begin{equation}\label{}
w_{x} = - 1 + \frac{\frac{2m}{1+3m} +  3m\Omega_{m0}}{1 - \Omega_{m0}} \qquad\qquad (a=1).
\end{equation}
Given that $|m|$ is small, this remains in the vicinity of $w_{x} = - 1$.
In the far future $w_{x}$ approaches
\begin{equation}\label{}
w_{x} \approx - 1 + \frac{2m}{1+3m}  \qquad\qquad (a\gg 1).
\end{equation}
The evolution of the effective EoS parameter $w_{x}$ is visualized in Fig.~2 for different values of $m$.
For small values of $a$ one has $w_{x}> 0$ but there will be a change to $w_{x}< 0$ well before the present epoch.
The effective energy density changes from the phantom regime to a later phase with $\rho_{x}> 0$.
Due to the sign change in $\rho_{x}$ the transition point $\rho_{x} = 0$ is accompanied by a singularity in $w_{x}$ at this point for $m$ positive.



\begin{figure}
\begin{center}
\includegraphics[width=0.45\textwidth]{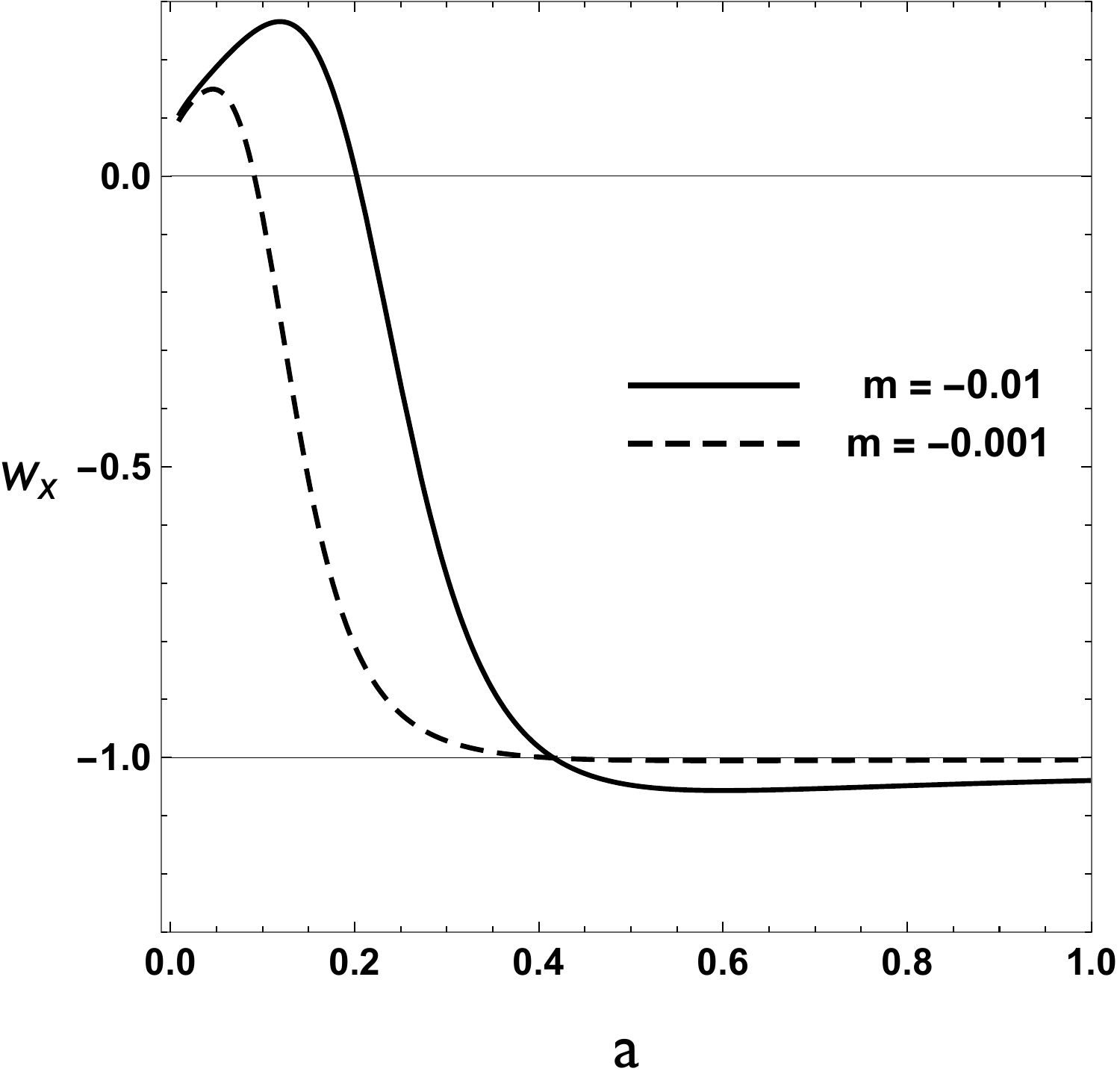}
\includegraphics[width=0.45\textwidth]{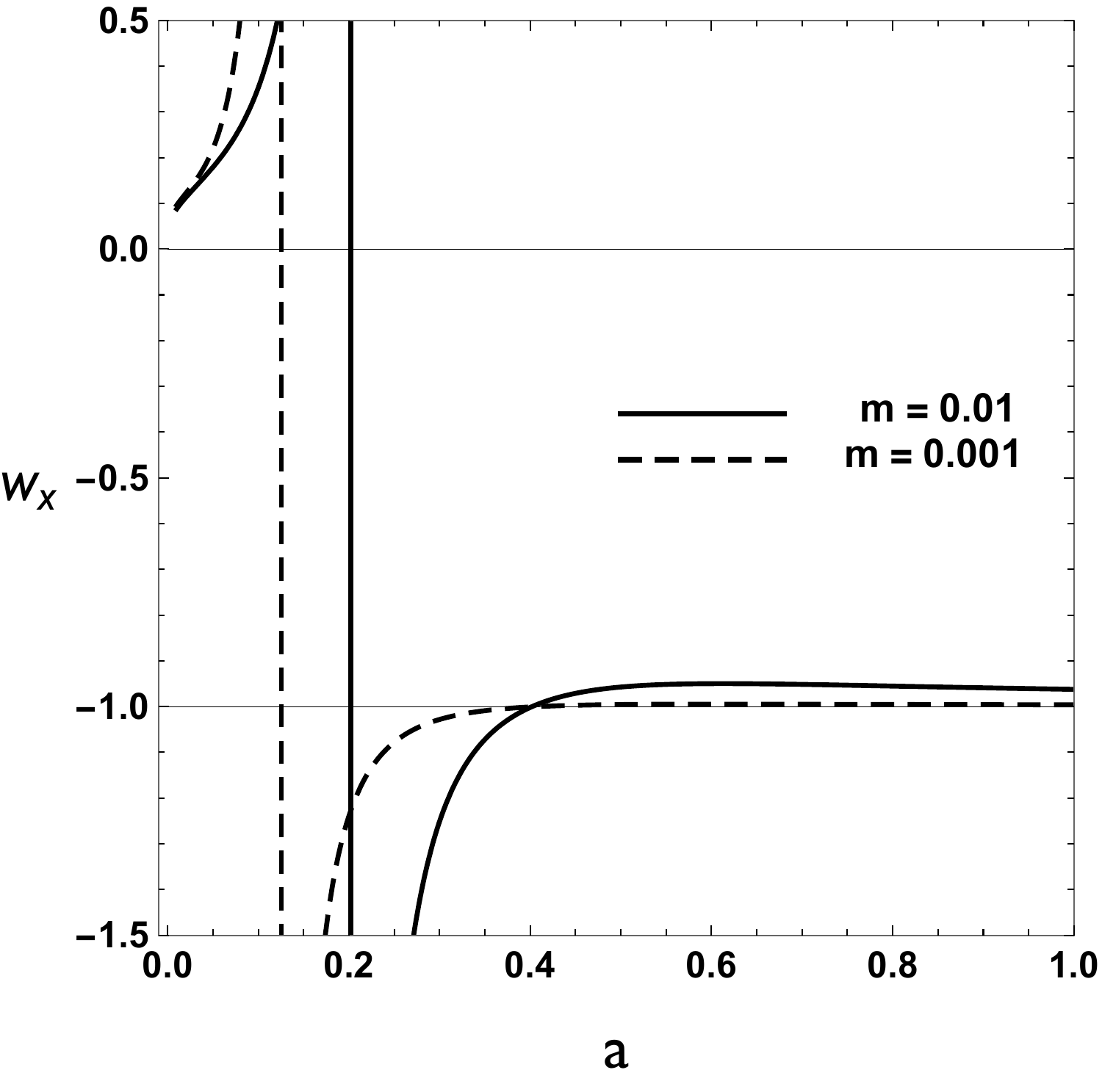}
\label{figOmxp}	
\caption{Effective EoS parameter as function of the scale factor for negative (left panel) and positive (right panel) values of $m$.}
\end{center}
\end{figure}

The result for the deceleration parameter $q = -1 - a\frac{H^{\prime}}{H}$ is
\begin{equation}\label{qJ}
q = \frac{1}{2}\frac{\frac{1-3m}{1+3m}A\Omega_{m0}a^{-3}\Phi^{-1}
- \frac{2}{1+3m}\left[1 - A\Omega_{m0}\right]\Phi}
{A\Omega_{m0}a^{-3}\Phi^{-1}
+ \left[1 - A\Omega_{m0}\right]\Phi}.
\end{equation}
As shown in Fig.~3, it changes from a high-redshift value
\begin{equation}\label{}
q\approx \frac{1}{2}\frac{1-3m}{1+3m} \qquad\qquad (a\ll 1),
\end{equation}
close to $\frac{1}{2}$, to the value close to $-1$,
\begin{equation}\label{}
q\approx -\frac{1}{1+3m} \qquad\qquad (a\gg 1),
\end{equation}
at late times.

\begin{figure}
\begin{center}
\includegraphics[width=0.45\textwidth]{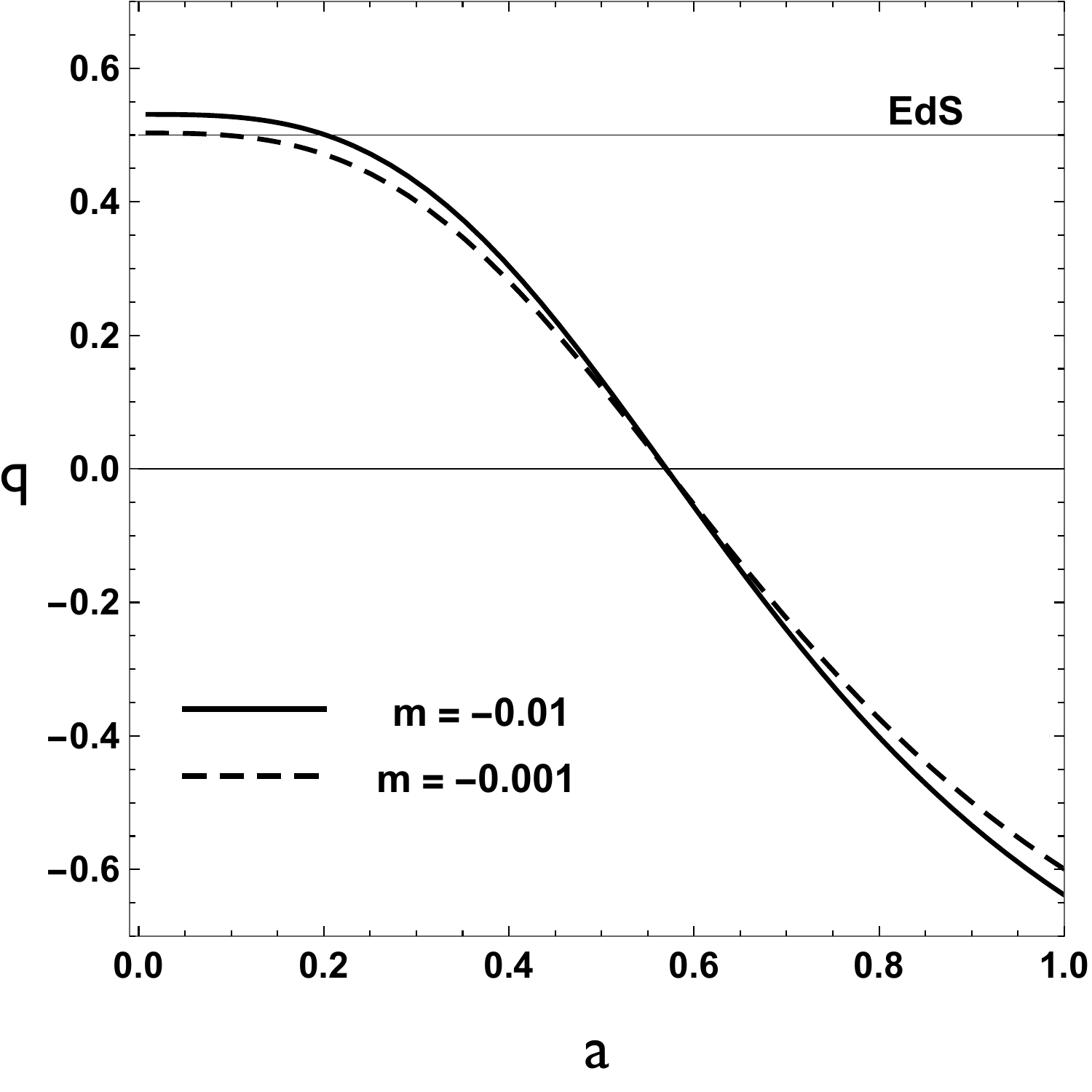}
\includegraphics[width=0.45\textwidth]{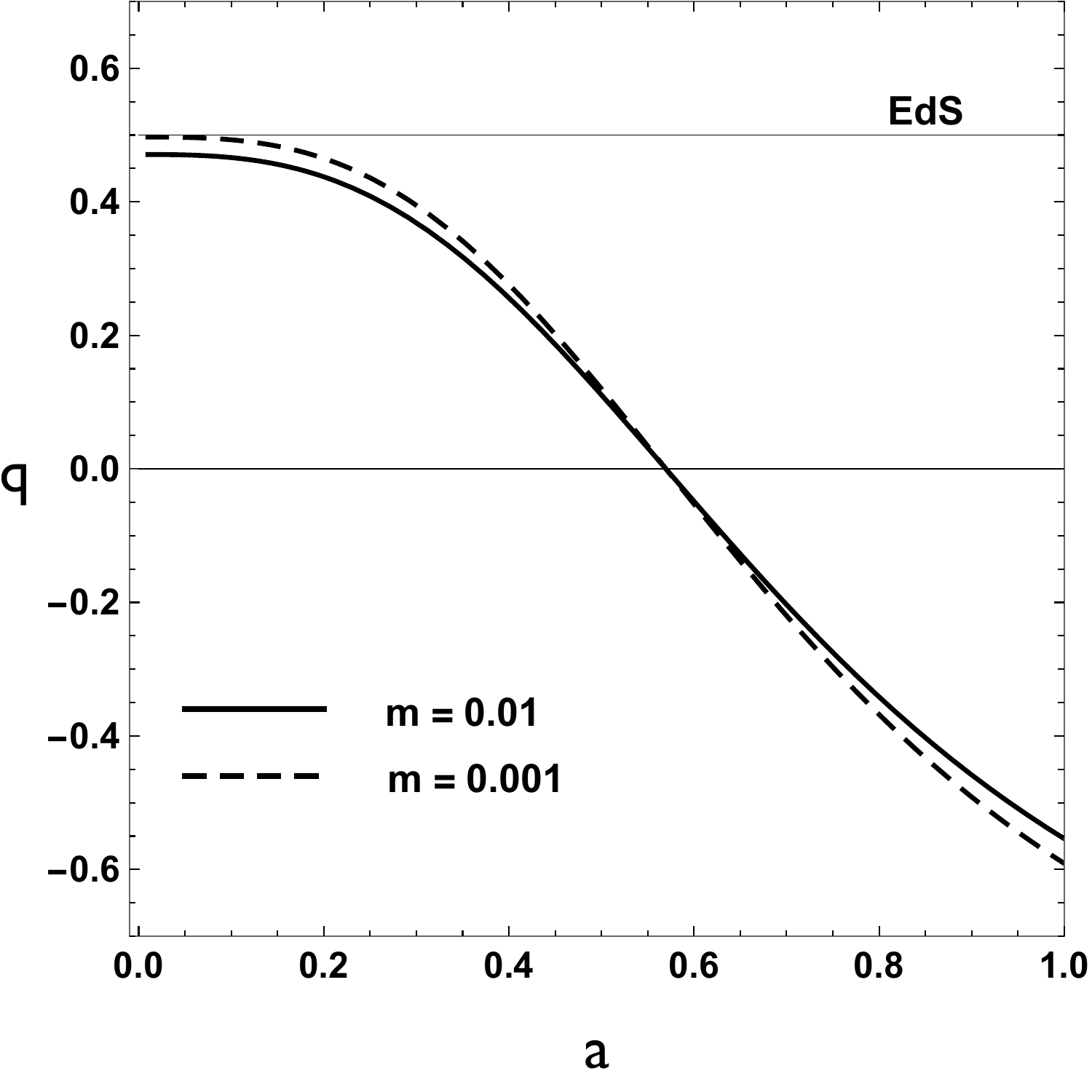}
\label{Fig5}	
\caption{Deceleration parameter for various negative (left panel) and positive (right panel) values of $m$.}
\end{center}
\end{figure}



The explicit expression for the pressure is
\begin{equation}\label{}
p_{x} = w_{eff}\rho_{x} = - \rho_{x}  + \frac{2m}{1+3m}\rho_{0}\left[1 - A\Omega_{m0}\right]\Phi +
\rho_{0}\Omega_{m0}a^{-3}\left[\frac{1+m}{1+3m}A\Phi^{-1} -1\right].
\end{equation}
Differentiating and using $\dot{\Phi} = - \frac{6m}{1+3m}H\Phi$
yields
\begin{equation}\label{}
\dot{p}_{x} = - \dot{\rho}_{x} - 3H \rho_{0}\left\{\left(\frac{2m}{1+3m}\right)^{2}\left[1 - A\Omega_{m0}\right]\Phi
+\Omega_{m0}a^{-3}\left[\left(\frac{1+m}{1+3m}\right)^{2}A\Phi^{-1} -1\right]
\right\}.
\end{equation}
Since
\begin{equation}\label{}
\dot{\rho_{x}} = -3H  \rho_{0}\left\{\frac{2m}{1+3m}\left[1 - A\Omega_{m0}\right]\Phi
+\Omega_{m0}a^{-3}\left[\frac{1+m}{1+3m}A\Phi^{-1} -1\right]
\right\},
\end{equation}
we find
\begin{equation}\label{}
\frac{\dot{p}_{x}}{\dot{\rho_{x}}} = -1 + \frac{\left(\frac{2m}{1+3m}\right)^{2}\left[1 - A\Omega_{m0}\right]\Phi
+\Omega_{m0}a^{-3}\left[\left(\frac{1+m}{1+3m}\right)^{2}A\Phi^{-1} -1\right]}
{\frac{2m}{1+3m}\left[1 - A\Omega_{m0}\right]\Phi
+\Omega_{m0}a^{-3}\left[\frac{1+m}{1+3m}A\Phi^{-1} -1\right]}.
\end{equation}
At high redshift this quantity (which corresponds to the adiabatic sound speed) is considerably smaller than 1,
it may even be close to zero.
The far-future limit is
\begin{equation}\label{dpxgg}
\frac{\dot{p}_{x}}{\dot{\rho_{x}}} \approx -1 + \frac{2m}{1+3m} \qquad\qquad (a\gg 1).
\end{equation}
While this does not seem to differ substantially from the standard model, the intermediate behavior does.
As visualized in Fig.~4, this quantity changes its sign at the points with $\dot{\rho}_{x} =0$ which implies a singularity in $\frac{\dot{p}_{x}}{\dot{\rho_{x}}}$. Recall that the energy density $\rho_{x}$ is an effective quantity which
 simulates DE but does not belong to the matter part of the field equations. It is of entirely geometric origin.



\begin{figure}
\begin{center}
\includegraphics[width=0.45\textwidth]{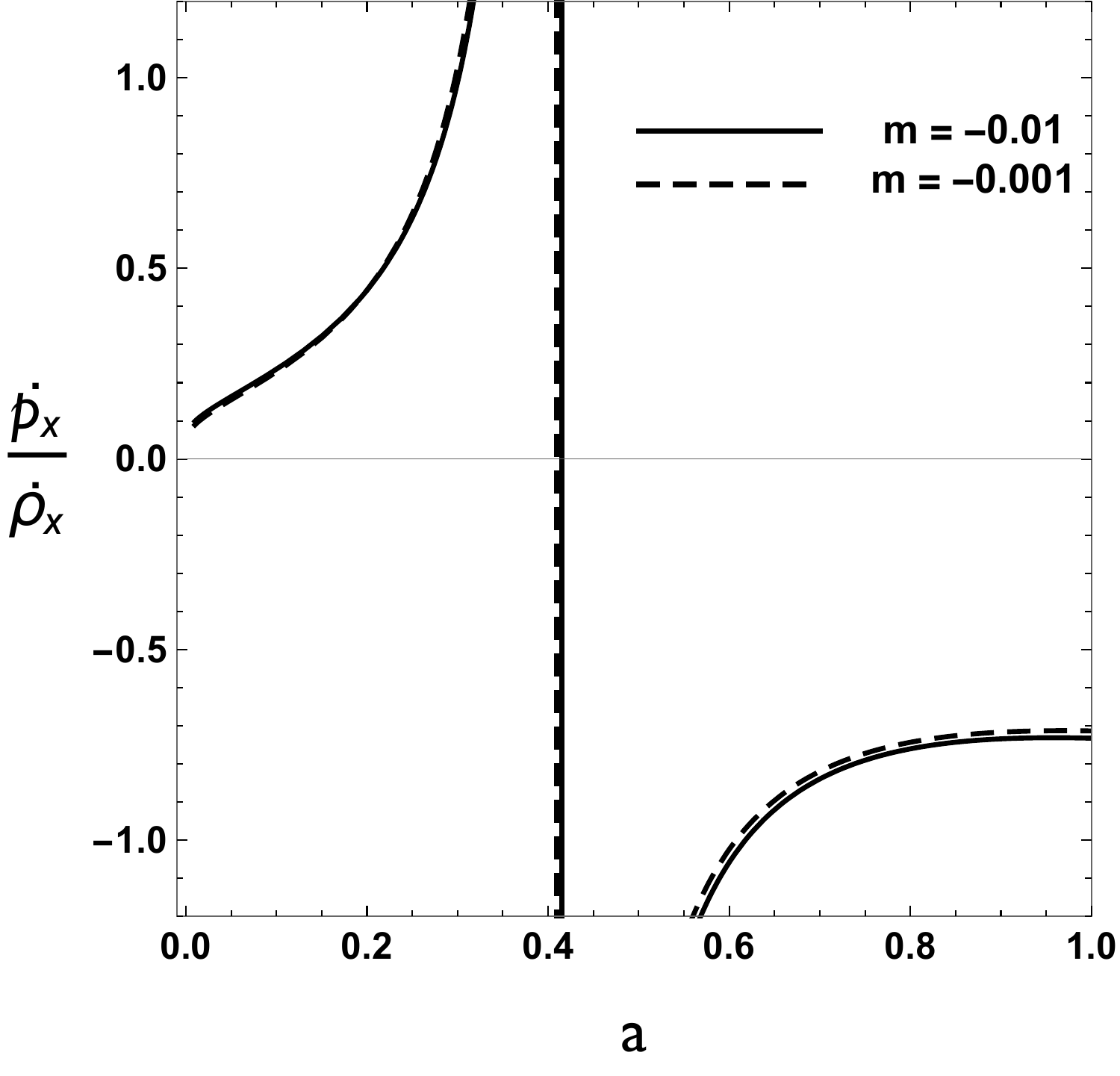}
\includegraphics[width=0.45\textwidth]{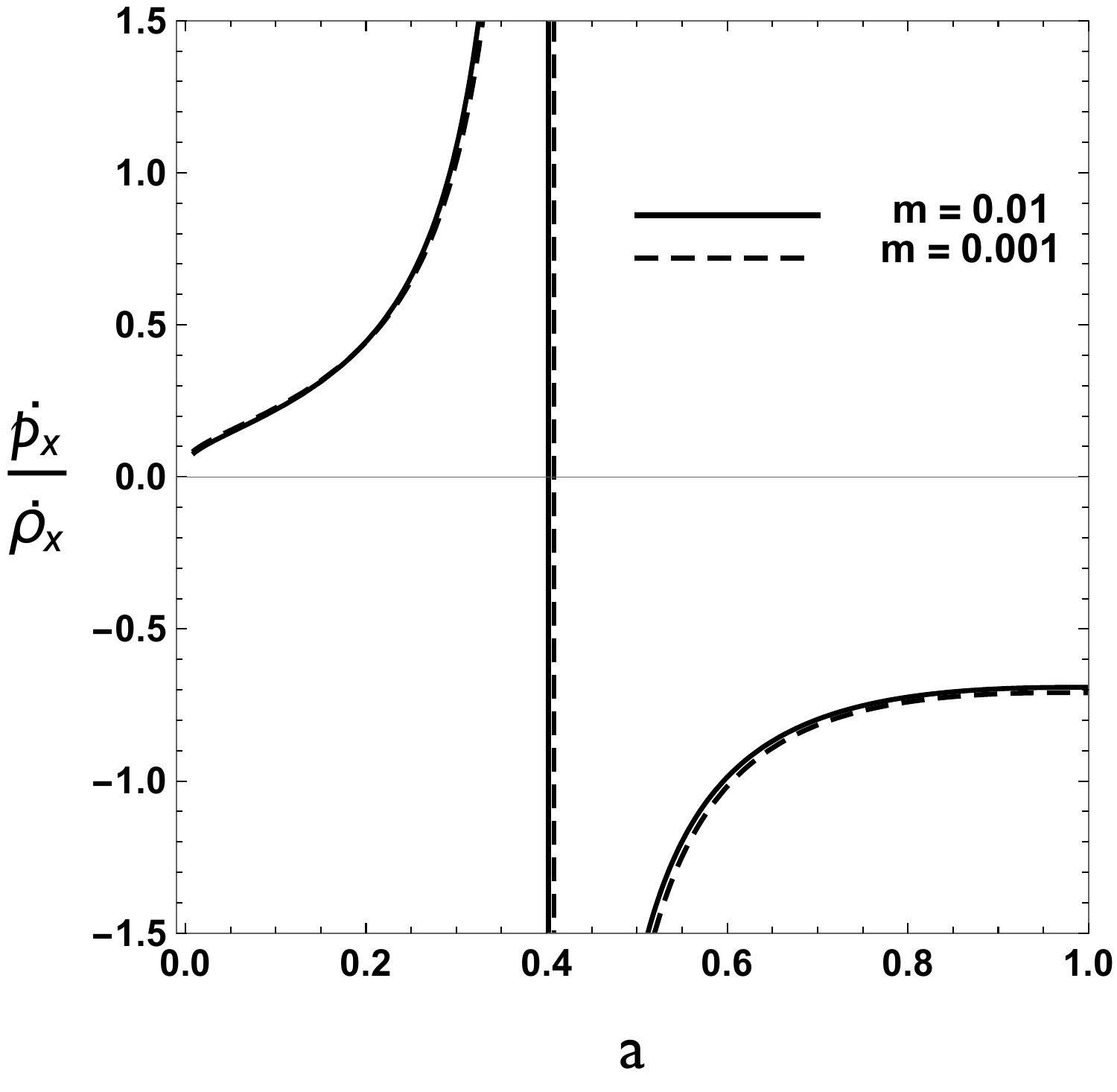}
\label{Fig8}	
\caption{Effective adiabatic sound speed for various negative (left panel) and positive (right panel) values of $m$.}
\end{center}
\end{figure}

We may define a total EoS parameter $w \equiv \frac{p}{\rho} = \frac{p_{x}}{\rho}$ which results in
\begin{equation}\label{}
w \equiv \frac{p}{\rho} = \frac{p_{x}}{\rho} = - \frac{1}{1+3m}
\frac{\left(1+m\right)\left[1 - A\Omega_{m0}\right]\Phi + 2m A\Phi^{-1}\Omega_{m0}a^{-3}}{\left[1 - A\Omega_{m0}\right]\Phi + A\Phi^{-1}\Omega_{m0}a^{-3}}.
\end{equation}
The limits are
\begin{equation}\label{}
w \approx -\frac{2m}{1+3m}\qquad\qquad (a\ll 1)
\end{equation}
and
\begin{equation}\label{}
w \approx -\frac{1+m}{1+3m}\qquad\qquad (a\gg 1),
\end{equation}
i.e., at high redshift $w$ is close to zero, in the far-future limit it is close to $-1$.
The total EoS parameter is visualized in Figs.~5. It is well behaved for all values of $m$.



\begin{figure}
\begin{center}
\includegraphics[width=0.45\textwidth]{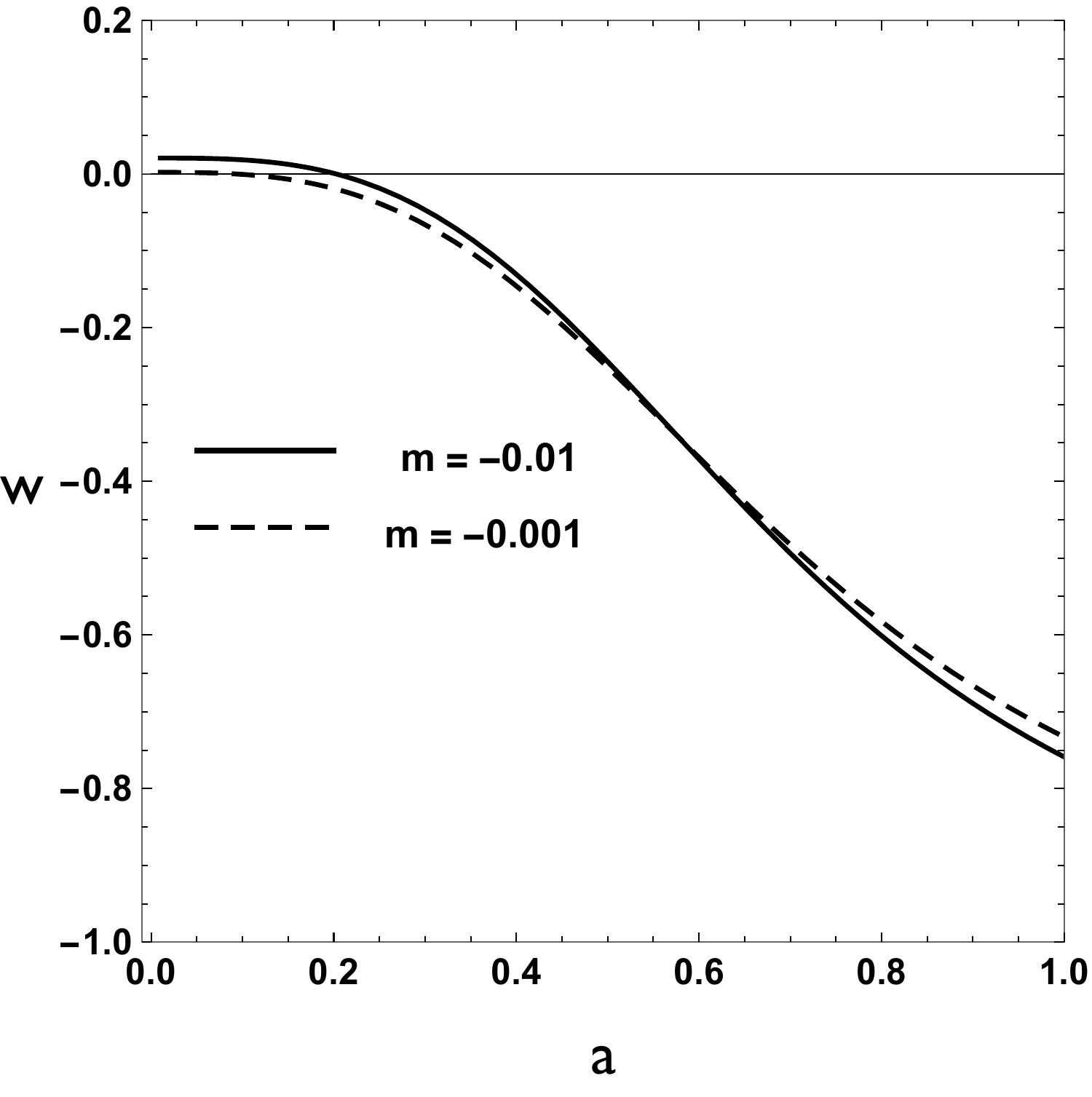}
\includegraphics[width=0.45\textwidth]{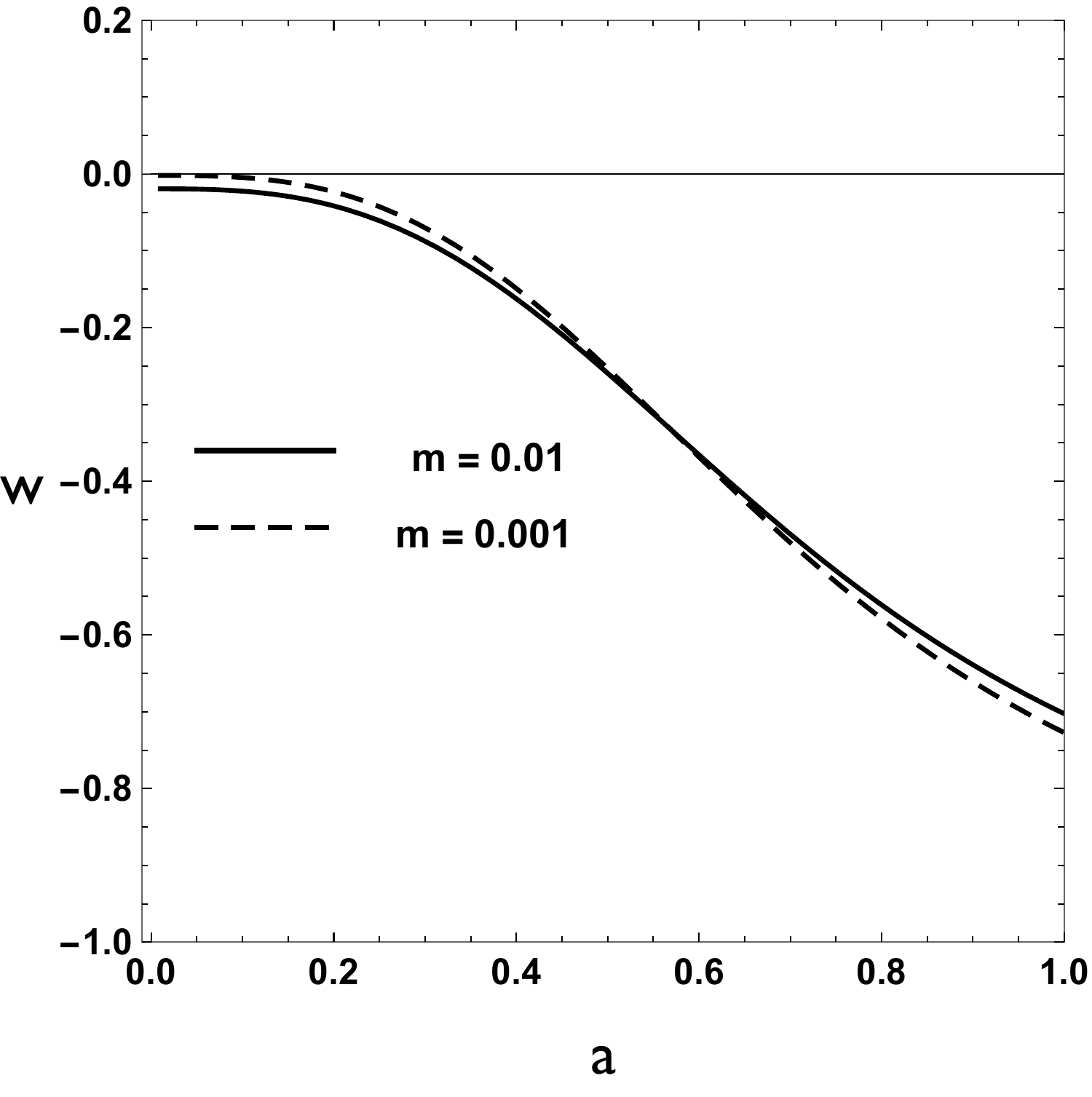}
\label{figOmxp}	
\caption{Total EoS parameter for various negative (left panel) and positive (right panel) values of $m$.}
\end{center}
\end{figure}

With
\begin{equation}\label{}
\frac{\rho + p}{\rho_{0}} = \frac{1+m}{1+3m}A\Phi^{-1}\Omega_{m0}a^{-3} + \frac{2m}{1+3m}\left[1 - A\Omega_{m0}\right]\Phi,
\end{equation}
\begin{equation}\label{}
\frac{\dot{\rho}}{\rho_{0}} = -3H \left\{\frac{1+m}{1+3m}A\Phi^{-1}\Omega_{m0}a^{-3} + \frac{2m}{1+3m}\left[1 - A\Omega_{m0}\right]\Phi\right\} = - 3H \frac{\rho + p}{\rho_{0}}
\end{equation}
and
\begin{equation}\label{}
\frac{\dot{p}}{\rho_{0}} = \frac{\dot{p}_{x}}{\rho_{0}}
= 3H \frac{2m\left(1+m\right)}{\left(1+3m\right)^{2}}\left[\left(1-A\Omega_{m0}\right)\Phi + \Omega_{m0}a^{-3}A\Phi^{-1}\right]
\end{equation}
we find
\begin{equation}\label{}
\frac{\dot{p}}{\dot{\rho}} = - \frac{2m\left(1+m\right)}{1+3m}
\frac{\left(1-A\Omega_{m0}\right)\Phi + \Omega_{m0}a^{-3}A\Phi^{-1}}{2m\left[1 - A\Omega_{m0}\right]\Phi
+ \left(1+m\right)A\Phi^{-1}\Omega_{m0}a^{-3}}.
\end{equation}
This effective adiabatic sound speed of the cosmic medium as a whole has the high-redshift limit
\begin{equation}\label{}
\frac{\dot{p}}{\dot{\rho}}\approx - \frac{2m}{1+3m}\qquad\qquad (a\ll 1),
\end{equation}
i.e., a very small value, and
\begin{equation}\label{}
\frac{\dot{p}}{\dot{\rho}}\approx - \frac{1+m}{1+3m}\qquad\qquad (a\gg 1),
\end{equation}
in the far future, a value close to $-1$.

While for $m>0$ the derivative $\dot{p}$ is always positive and  $\dot{\rho}$ is always negative, the quantity $\frac{\dot{p}}{\dot{\rho}}$ remains negative in the entire range.
For $m<0$, on the other hand, $\dot{p}$ is always negative but $\dot{\rho}$ may change its sign for sufficiently large  values of $a$, which gives rise to a divergency in  $\frac{\dot{p}}{\dot{\rho}}$ at the critical value
\begin{equation}\label{as}
a_{s} = \left[\frac{\left(1-|m|\right)A\Omega_{m0}}{2|m|\left(1-A\Omega_{m0}\right)}\right]^{\frac{1}{3}\frac{1-3|m|}{1+|m|}}\qquad (m<0).
\end{equation}
For $a<a_{s}$ the ratio $\frac{\dot{p}}{\dot{\rho}}$  is positive, for $a>a_{s}$  it is negative since $\rho$ starts to grow at $a=a_{s}$. This is consistent with the far-future limit in (\ref{rhofut}).
Except for the rather large deviation from the standard model with $m=-0.1$, these singularities occur in the distant future.
Figs.~6 shows the dependence of $\frac{\dot{p}}{\dot{\rho}}$ on the scale factor for various values of m.

We finish this section with a simplified stability check of the solution for $\Phi$, leaving a true perturbation analysis to be the subject of a subsequent paper.
We introduce a decomposition
\begin{equation}\label{Phisplit}
\Phi = \Phi_{b} + \Phi_{1}
\end{equation}
with our ``background" solution $\Phi_{b}$ and a (homogeneous) perturbation $\Phi_{1}$ according to
\begin{equation}\label{}
\Phi_{b} = a^{-\frac{6m}{1+3m}}, \qquad \Phi_{1} \ll \Phi_{b}.
\end{equation}
Linearizing yields
\begin{equation}\label{}
\Phi^{\frac{1+3m}{2m}} = \left(\Phi_{b} + \Phi_{1}\right)^{\frac{1+3m}{2m}}
\approx \Phi_{b}^{\frac{1+3m}{2m}} + \frac{1+3m}{2m}\Phi_{b}^{\frac{1+m}{2m}}\Phi_{1}
\end{equation}
and
\begin{equation}\label{}
\Phi^{2} = \left(\Phi_{b} + \Phi_{1}\right)^{2}
\approx \Phi_{b}^{2} + 2\Phi_{b}\Phi_{1}.
\end{equation}
Introducing  these decompositions into (\ref{equeff}) provides us, at first order,  with an equation for $\Phi_{1}$,
\begin{equation}\label{}
\frac{d^{2}\Phi_{1}}{dt^{2}} + 3H\frac{d\Phi_{1}}{dt}
+ 9H_{0}^{2} \left[\frac{1}{2}\frac{1+m}{1+3m} A\Omega_{m0}\Phi_{b}^{\frac{1+m}{2m}}
+ \frac{4m}{\left(1+3m\right)^{2}}\left(1 - A\Omega_{m0}\right)\Phi_{b}\right]\Phi_{1} =0.
\end{equation}
This is an equation for a damped harmonic oscillator with a time-dependent frequency $\omega$, given by
\begin{equation}\label{om2}
\omega^{2} = 9H_{0}^{2} \left[\frac{1}{2}\frac{1+m}{1+3m} A\Omega_{m0}\Phi_{b}^{\frac{1+m}{2m}}
+ \frac{4m}{\left(1+3m\right)^{2}}\left(1 - A\Omega_{m0}\right)\Phi_{b}\right].
\end{equation}
The perturbation $\Phi_{1}$ is bounded, i.e. there is no instability for any $\omega^{2} > 0$. For positive values of $m$ this is always guaranteed. For $m<0$ the second term in the square brackets had to dominate the first one to violate the condition $\omega^{2} > 0$. But since  we imposed $|m|\ll 1$ this may happen only under the extreme condition
$a>a_{cr}$ where $a_{cr}$ is a critical value, given by
\begin{equation}\label{}
a_{cr}^{3\frac{1+|m|}{1-3|m|}} = \frac{1-3|m|}{8|m|}\left(1-|m|\right)\frac{A\Omega_{m0}}{1 - A\Omega_{m0}}.
\end{equation}
It follows that an instability may occur at the most
in the far-future limit $a> a_{cr}$ where $a_{cr}\gg 1$ due to the inverse dependence on $|m|$. This value of $a_{cr}$ is of the same order as the previously derived
critical scale factor (\ref{as}).
In Fig.~7 we depict the dependence (\ref{om2}) of $\omega^{2}$ on $m$ for the present value $\Phi_{b} =1$ (solid line) and  for a constant value $\Phi_{b} <1$ (broken line) which corresponds to a scale factor in the past, i.e. $a<1$, for $m<0$.
The gray strip represents the unstable part $\omega^{2}<0$.
 The observationally allowed
region for $m$ (see the statistical analysis of the following section) lies entirely in the stable range, even for the admitted negative $m$-values. Towards the past, the range of admissible negative $m$-values increases since the intersection of the broken line with $\omega^{2} = 0$ is shifted to the left.

With the help of the decomposition  (\ref{Phisplit}) the Hubble rate (\ref{H2Phi}) may be split into a background part
$H_{b}$ and linear, homogeneous perturbations $H_{1}$ about this background as well.
Since with $a^{-3} = \Phi^{\frac{1+3m}{2m}}$ the time evolution of the Hubble rate (\ref{H2Phi}) is entirely determined by the dynamics of $\Phi$, we find $H_{1} \propto \Phi_{1}$, i.e., the perturbed Hubble rate inherits the stability properties of the scalar field perturbations.



\begin{figure}
\begin{center}
\includegraphics[width=0.45\textwidth]{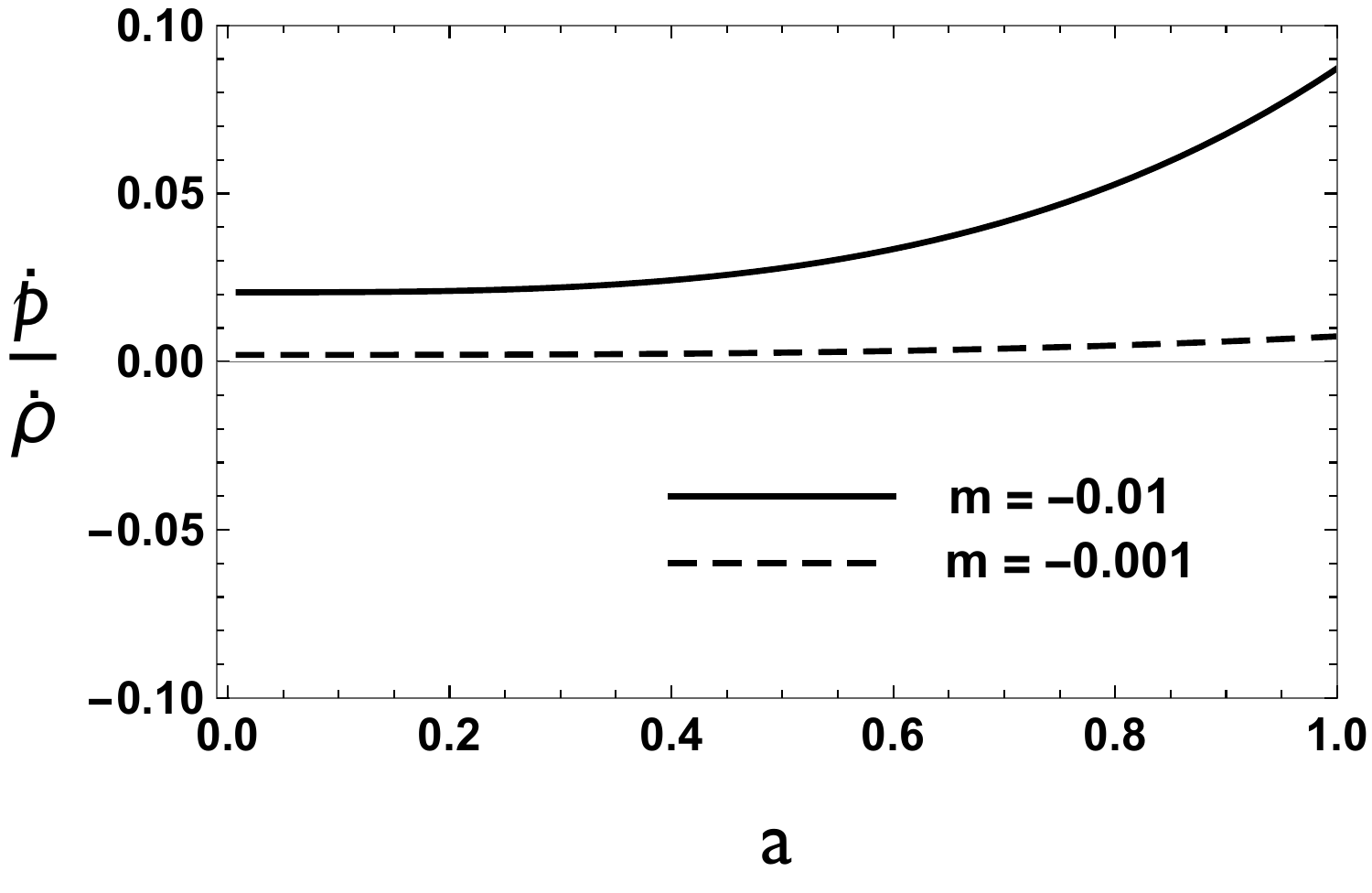}
\includegraphics[width=0.45\textwidth]{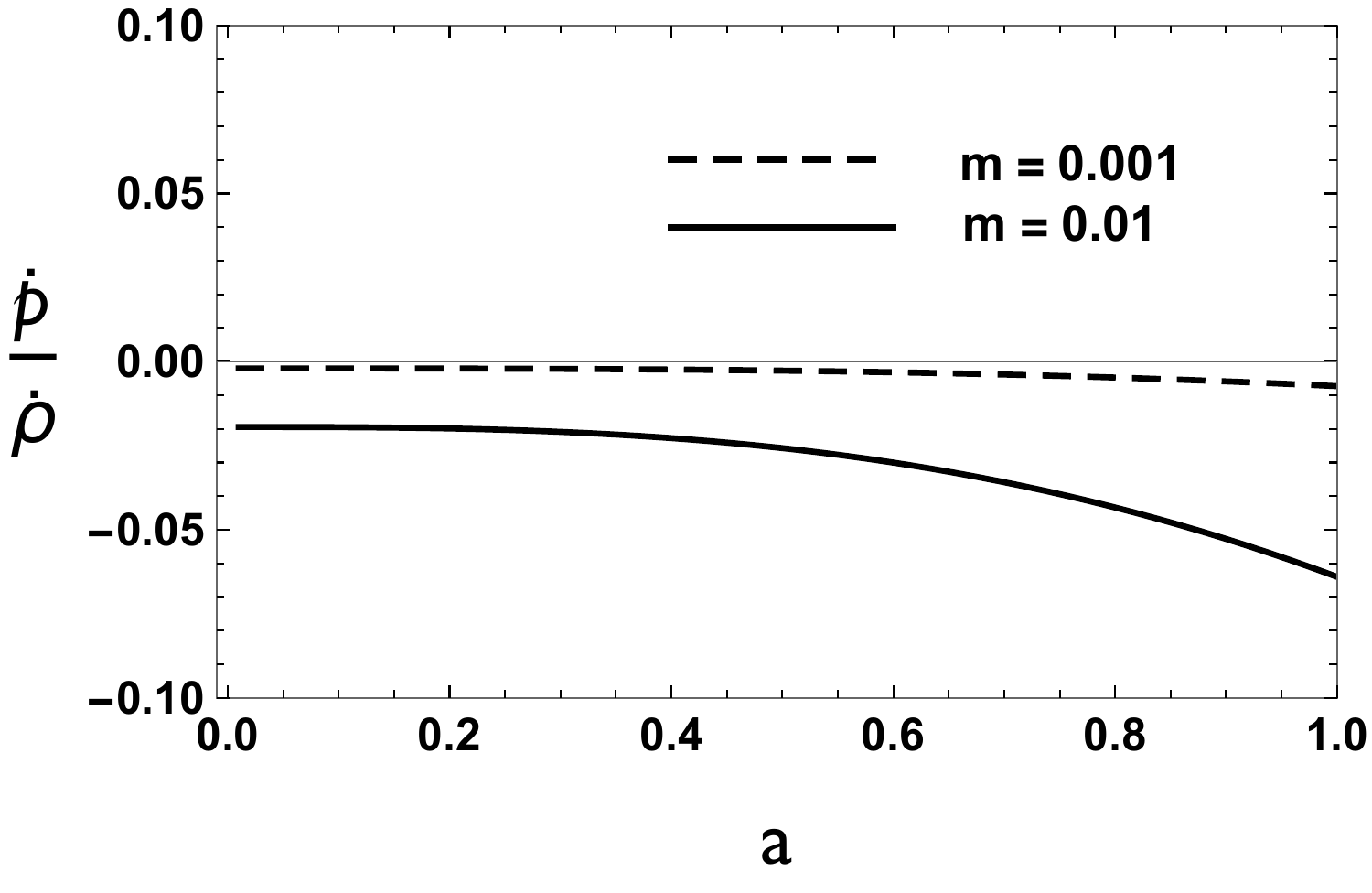}
\label{Fig12}	
\caption{Scale-factor dependence of the total effective adiabatic sound speed for various negative (left panel) and positive (right panel) values of $m$.}
\end{center}
\end{figure}

\begin{figure}
\begin{center}
\includegraphics[width=0.55\textwidth]{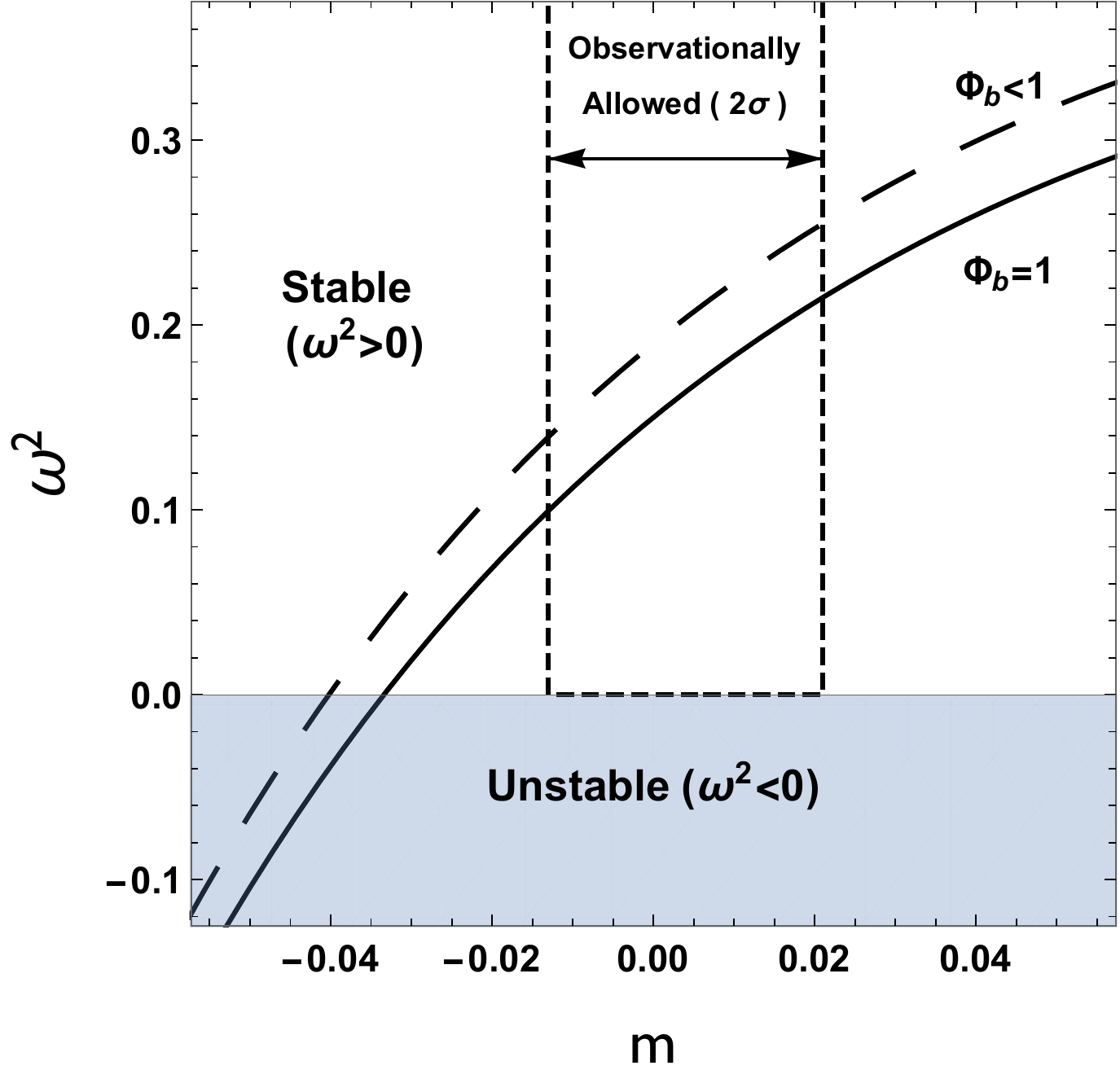}
\label{figOmxp}	
\caption{
Dependence of $\omega^{2}$ on $m$ for the present value $\Phi_{b} =1$ (solid line) and  for a constant value $\Phi_{b} <1$ (broken line) which, for $m<0$,  corresponds to a scale factor in the past, i.e. $a<1$. The instability
range ($m$-values left to the intersection of the $\Phi_{b}$ curve with the axis $\omega^{2} =0$ ) shrinks towards the past, since the intersection point moves to the left for $\Phi_{b} <1$. The grey stripe marks the instability region. The observationally allowed values lie entirely in the stable part of the $\omega^{2}$ - $m$ plane.
For $m>$ stability is always guaranteed. }
\end{center}
\end{figure}

\section{Statistical (Observational) analysis}
\label{observations}

The purpose of this section is to test the viability of the background expansion predicted by (\ref{H2}), equivalent to  (\ref{H2Phi}), based on the available observations. A particular interest here is to obtain constraints on the parameter $m$ which is responsable to dictating deviations of our model from $\Lambda$CDM.

In order to compare (\ref{H2}) (or (\ref{H2Phi})) with the data we constrain the model parameters with the following observational data sets.

\noindent {\bf Supernovae}: First, we use Supernovae data from the JLA compilation \cite{JLA}. We will use the binned data set provided by reference \cite{JLA} with the corresponding covariance matrix $\bf{C}$. This test is based on the observed distance modulus $\mu_{obs}(z)$ of each binned SN Ia data at a certain redshift $z$,
\begin{equation}\label{muth}
\mu_{th}(z)=25+5 log_{10}\frac{d_L(z)}{Mpc},
\end{equation}
where the luminosity distance, in a spatially flat FLRW metric, is given by the formula
\begin{equation}
d_L(z)=c(1+z)\int^{z}_0\frac{dz^{\prime}}{H(z^{\prime})}.
\end{equation}
Knowledge of the Hubble expansion rate allows us to compute the predicted theoretical value $\mu_{th}(z_i)$ for a given redshift $z_i$. The binned JLA data contains 31 data-points. For the JLA Supernova sample the $\chi^2$ function is constructed according to
\begin{equation}
\chi^{2}_{SN}= \left(\bf{\mu_{th}(z)-\mu_{obs}(z)}\right)^{\dagger} C^{-1} \left(\bf{\mu_{th}(z)-\mu_{obs}(z)}\right).
\end{equation}

\noindent{\bf Differential Age}:
A second observational source comes from the evaluation of differential age data of old galaxies that have evolved passively \cite{Ji1, Ji2, Hz}. Indeed, the expansion rate is defined as
\begin{equation}
H(z)=-\frac{1}{1+z}\frac{dz}{dt}.
\end{equation}
Since spectroscopic redshifts of galaxies are known with very high accuracy, one just needs a differential measurement of time $dt$ at a given redshift interval in order to obtain values for H(z). The data used in this work consist of $28$ data points listed in \cite{Farooq}, but previously compiled in \cite{moresco}.

\noindent {\bf Baryon Acoustic Oscillations}:
The baryon acoustic oscillation (BAO) scale is calculated via
\begin{equation}
D_V(z)=\left[(1+z)^2 D_A^2(z)\frac{c z}{H(z)}\right]^{1/3},
\end{equation}
where $D_A(z)$ is the angular-diameter distance. The values for $D_V$ have been reported in the literature by several galaxy surveys. In our analysis we use data at $z=0.2$ and $z=0.35$ from the SDSS \cite{sdss}, data at $z=0.44,0.6$ and $0.73$ from the WiggleZ \cite{wiggle} and one data point at $z=0.106$ from the 6dFGRS \cite{6dfgrs} surveys.

For our statistical analysis we construct the chi-square function of each sample
\begin{equation}
\chi^2=\sum^{N}_{i=1}\frac{\left(f^{th}(z_i)-f^{obs}(z_i)\right)^2}{\sigma^2_i},
\end{equation}
where $f=\left(H,D_V\right)$ for the H(z) and BAO datasets, respectively. The number of data points $\left\{z_i,f^{obs}(z_i)\right\}$ in each set is, respectively, $N_{H}$ and $N_{BAO}$
whereas $\sigma_i$ is the observational error associated to each observation $f^{obs}$ and $f^{th}$ is the theoretical value predicted by the scalar-tensor model.

Adding up information from all data samples we can construct the total chi-square function as
\begin{equation}\label{chitotal}
\chi^2_{Total}=\chi^2_{SN}+\chi^2_{H}+\chi^2_{BAO}.
\end{equation}

We consider the expansion rate (\ref{H2}) as a model with three free parameters, $H_0, \Omega_{m0}$ and $m$. Our main interest is in constraints on the latter.

We will fix two hypersurfaces for the parameter $H_0$: the Planck prior $H_0=68.7\mathrm{Km/s/Mpc}$ \cite{Planck} and the recent determination from the Hubble Space Telescope (HST) $H_0=73.2 \mathrm{Km/s/Mpc}$ \cite{RiessHST}. These priors on $H_0$ allows us to span the $\Omega_{m0}$ x $m$ plane. This can be seen in the left and central panels in Fig.~8 where the $1 \sigma$ and $2 \sigma$ confidence regions are shown in red for the H(z) data, blue for the Supernovae data and green for the BAO data, respectively, for both $H_0$ priors. The combined contours obtained from the total chi-square function (\ref{chitotal}) are given by the solid black lines.
From these plots it is clear that the preferred $m$-values depend on the $H_0$ prior. With the Planck prior positive values of $m$ are preferred, while the HST prior results in a preference for negative $m$.

It is worth noting that in both two-dimensional plots the $\Lambda$CDM model, expressed by the horizontal line at $m=0$, lies outside the $2\sigma$ confidence level region of the total ($\chi^2_{Total}$) function.

\begin{figure}
\begin{center}
\includegraphics[width=0.3\textwidth]{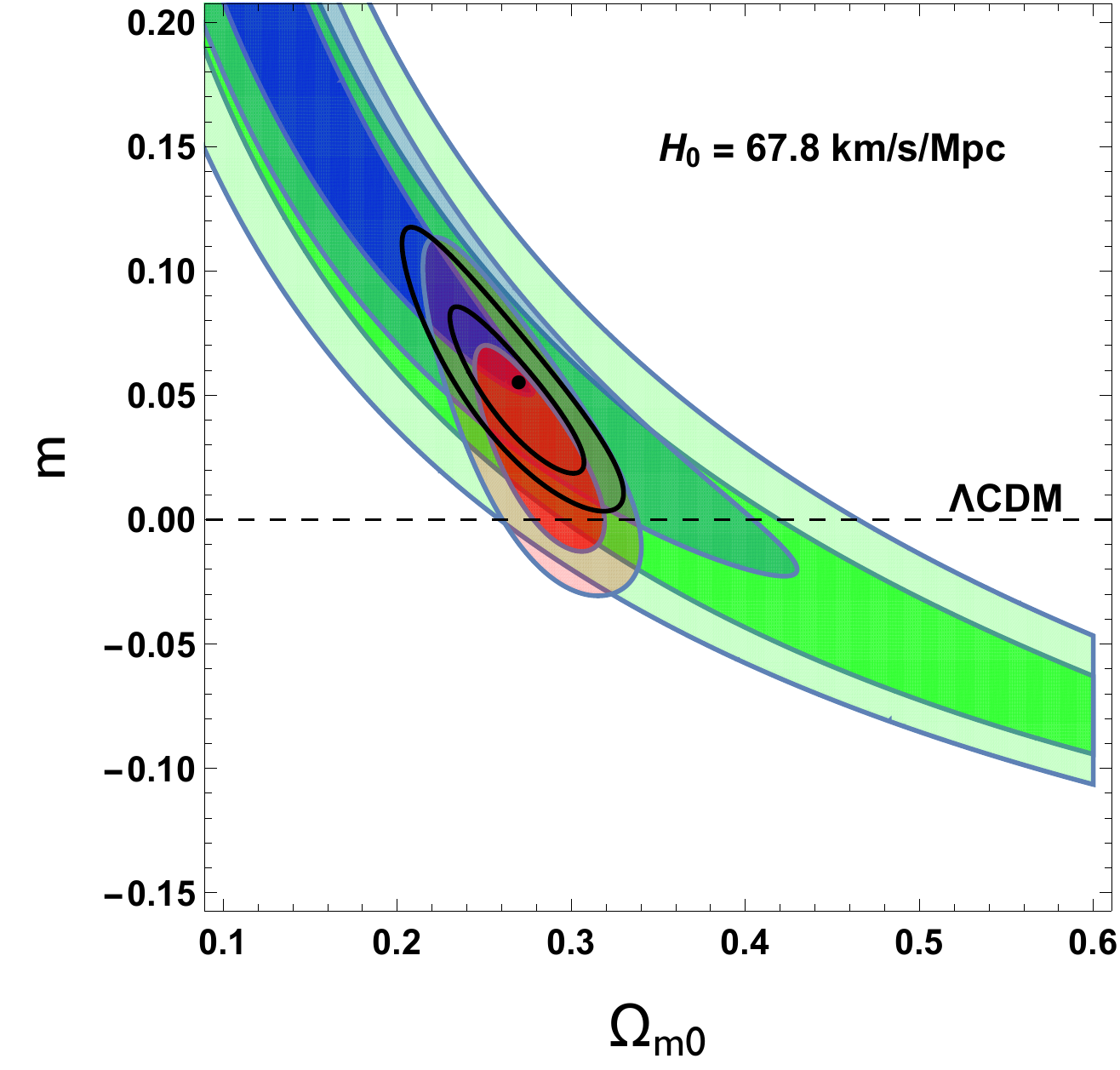}
\includegraphics[width=0.3\textwidth]{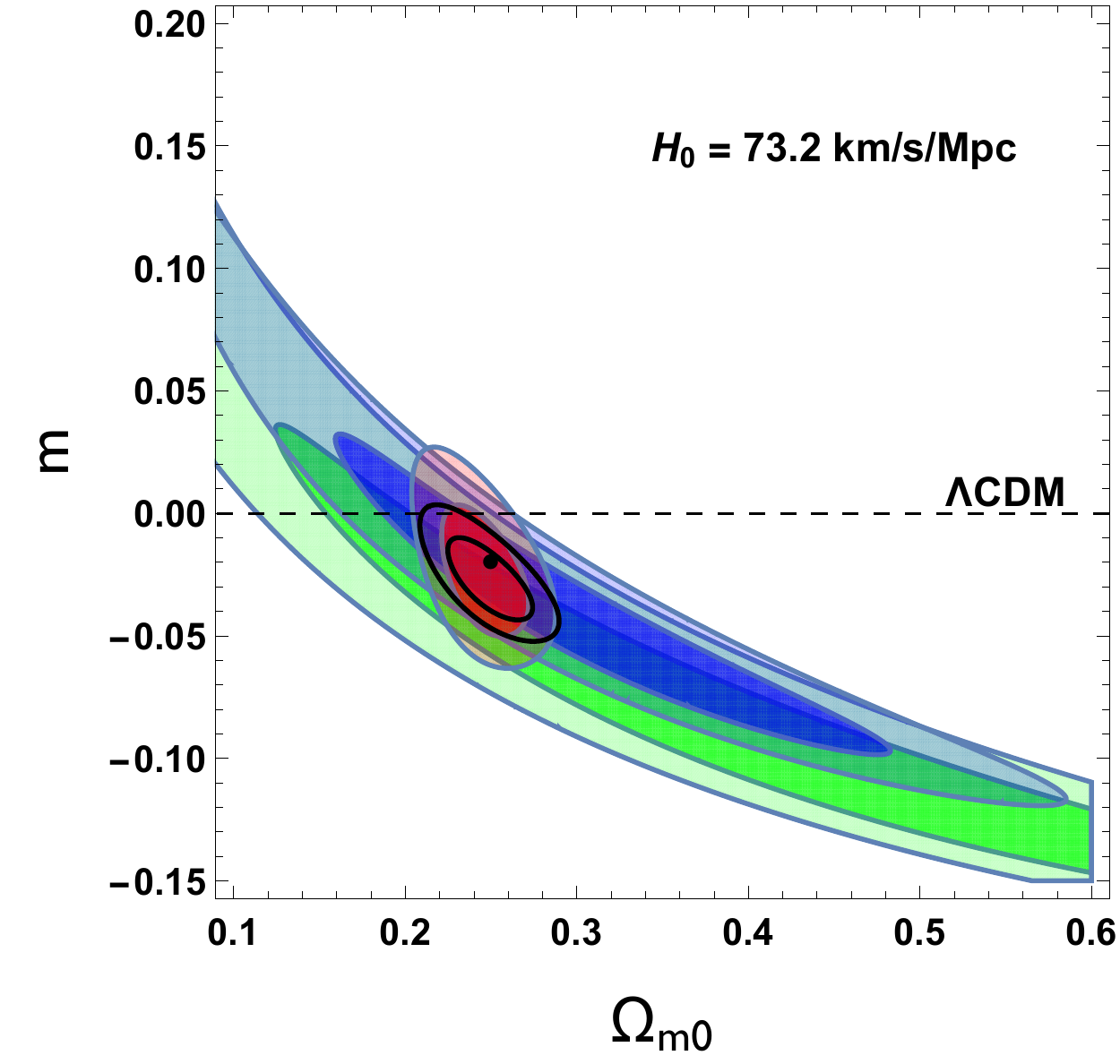}
\includegraphics[width=0.3\textwidth]{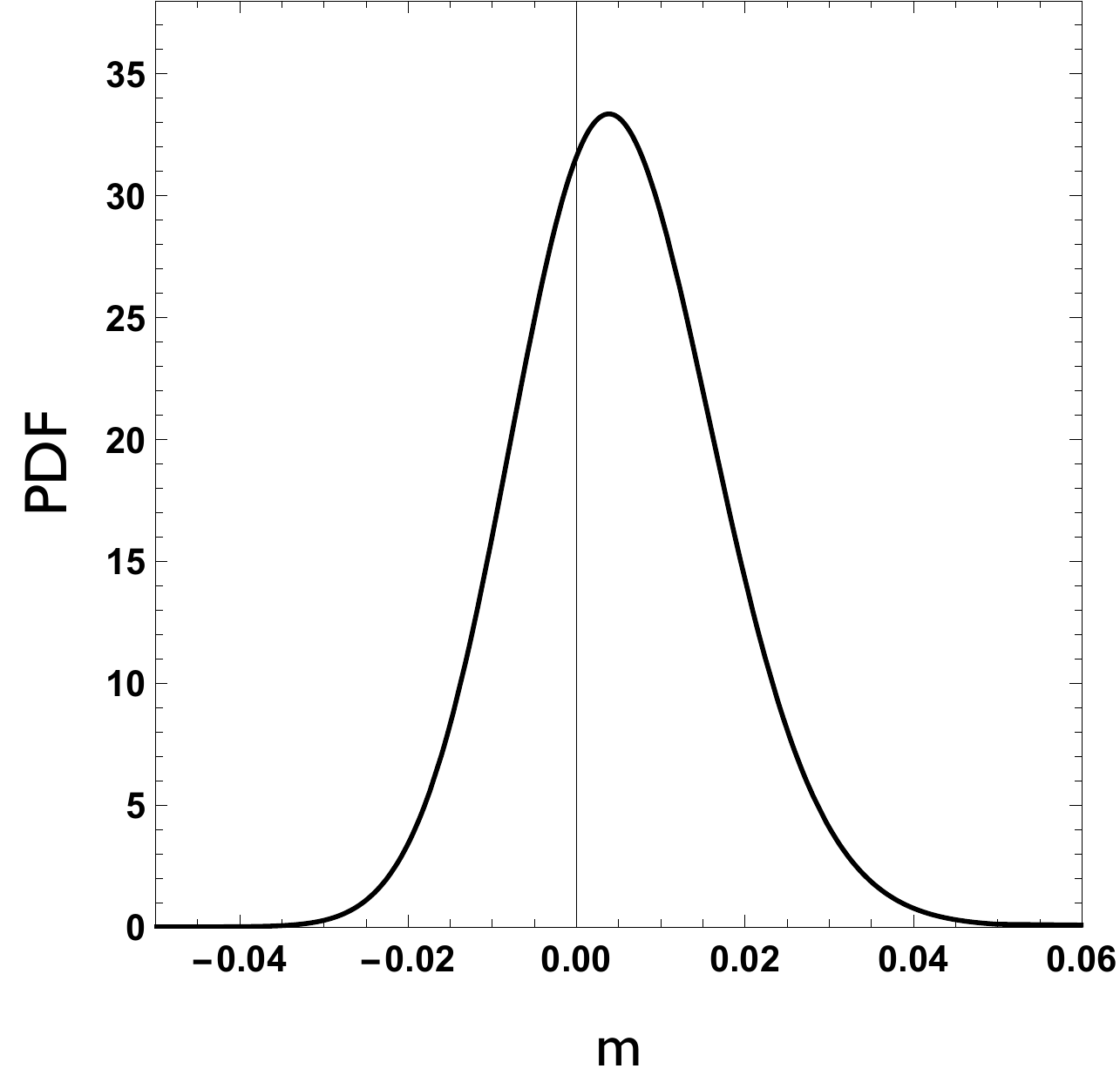}
\label{fig1}	
\caption{Observational constraints on the model parameters. In the left panel we fix the hypersurface with the Planck prior $H_0=67.8 \mathrm{km/s/Mpc}$ and display 1$\sigma$ and 2$\sigma$ confidence level contours in the plane $\Omega_{m0}$ x $m$. The central panel displays the same contours with the recent HST prior $H_0=73.2 \mathrm{km/s/Mpc}$. H(z) data only is represented by red regions, BAO data in green and Supernovae data in blue. The contours obtained from the total $\chi^2$ are denoted by the black solid lines. In the right panel we display the one-dimensional probability distribution function (PDF) for the parameter $m$ after marginalization over $H_0$ and $\Omega_{m0}$.}
\end{center}
\end{figure}

In order to obtain specific constraints on $m$ we apply Bayesian statistical analysis. With this procedure we obtain a one-dimensional probability distribution function (PDF) after marginalizing the likelihood function
\begin{equation}
\mathcal{L}= A e^{ -\chi^2_{Total}(H_0, \Omega_{m0}, m)/2}
\end{equation}
over the parameters $H_0$ and $\Omega_{m0}$.

The resulting PDF for the parameter $m$ is shown in the right panel of Fig. 8. Although positive values for the parameter $m$ are slightly preferred, the peak of the distribution has been obtained at $m= 0.004^{+0.011 (1\sigma) \,\, +0.017 (2\sigma)}_{-0.011 (1\sigma)\,\, -0.017 (2\sigma)}$ which indicates the agreement of the model with the $\Lambda$CDM model.


\section{Discussion}
\label{conclusions}

We have established a class of scalar-tensor-theory-based cosmological models which are simple extensions of the
$\Lambda$CDM model. The background dynamics of this class has been discussed in detail. Our main result is an analytic expression (\ref{H2Phi}) for the Hubble rate which explicitly quantifies the difference to the standard model through a constant parameter $m$ which determines the dynamics of the scalar field $\Phi$.
The solution for $\Phi$ is a consequence of the solution of the macroscopic fluid dynamics.
It corresponds to a scalar-field dynamics (\ref{eqPhieff}) with an effective potential given by (\ref{Ueff}).
We identified a geometric equivalent of the DE component of the standard model. The corresponding effective energy density may be positive or negative, including a transition between both regimes. The effective EoS parameter of this component diverges at the transition point but the overall dynamics is well behaved.
A similar comment holds for the effective adiabatic sound speed square of the geometrical DE.
Such behavior is unavoidable in any model in which the energy density of an DE equivalent and its time derivative change their signs during the cosmic evolution.
This seems to limit the usefulness of an effective fluid description of parts of the geometrical sector and may cause computational problems.
But since the overall dynamics and the dynamics of the matter component are smooth, the mentioned apparently unwanted features do not really jeopardize the model. They just demonstrate the fact that the fluid picture for parts of the geometry is a formal description which may well differ from that of a real fluid.

Our tests of the model parameter $m$ with data from SNIa, $H(z)$ and BAO constrain $m$ to values very close
to the $\Lambda$CDM value $m=0$ with a slight preference for positive values. We expect the existence of the analytic background solution (\ref{H2Phi}) to be useful for future investigation of the perturbation dynamics of this scalar-tensor extension of the $\Lambda$CDM model.
Even a very small non-vanishing value of $|m|$ will certainly modify the standard scenario of structure formation.\\
\ \\
\noindent
{\bf Acknowledgement:} Financial support by  CNPq, CAPES and FAPES is gratefully acknowledged.

\end{document}